\documentclass[a4paper,11pt]{article}
\pdfoutput=1

\usepackage{jheppub1}
\usepackage[T1]{fontenc}
\usepackage{amssymb}
\usepackage{graphicx}
\usepackage{amsmath}
\usepackage{hyperref}
\usepackage{tensor}
\usepackage{epstopdf}
\usepackage{extarrows}
\usepackage{verbatim}
\usepackage{subfigure}
\usepackage{float}
\usepackage{booktabs}  
\usepackage{hhline} 
\usepackage{tikz}

\newcommand{\td}{\text{d}}

\begin{document}
	
	\title{On Charged Black Holes in Einstein-Weyl-Maxwell Theory}
	
	\author{Ze Li and Hai-Shan Liu}
	\emailAdd{lize@tju.edu.cn}
	\emailAdd{hsliu.zju@gmail.com}
	\affiliation{Center for Joint Quantum Studies and Department of Physics, School of Science, Tianjin University, Yaguan Road 135, Jinnan District, 300350 Tianjin, P.~R.~China}
	
	\abstract{
		There exist two branches of static and spherically symmetric black hole solutions in Einstein-Weyl theory: one is the Schwarzschild black hole, and the other is a numerically constructed black hole that bifurcates from the Schwarzschild solution. Similarly, there are two branches of charged black holes in Einstein-Weyl-Maxwell theory. We have uncovered the relationships between the charged black holes and the neutral ones. The two charged black holes branch out from the bifurcation point of the neutral ones once charge is added.
		We found that one of the charged black holes is entirely different from the Reissner-Nordström (RN) black hole, while the other is similar to the RN black hole. In particular, the RN-like black hole approaches the RN black hole as the charge increases. We calculated the charge-to-mass ratio of the RN-like charged black hole in the near-extremal limit, and the value is less than that of the extremal RN black hole.	Since we consider the higher-derivative Weyl square term as part of the classical gravity theory, rather than as a quantum effect, our result sets a lower bound on the charge-to-mass ratio in the context of the Weak Gravity Conjecture.

	}
	
	% Uncomment for PACS numbers
	%\pacs{00.00, 20.00, 42.10}
	%
	% Uncomment for keywords
	%\vspace{2pc}
	%\noindent{\it Keywords}: XXXXXX, YYYYYYYY, ZZZZZZZZZ
	%
	% Uncomment for Submitted to journal title message
	%\submitto{\JPA}
	%
	% Uncomment if a separate title page is required
	\maketitle
	%
	% For two-column output uncomment the next line and choose [10pt] rather than [12pt] in the \documentclass declaration
	%\ioptwocol
	%
	%\tableofcontents
	
\section{Introduction}\label{section 1}
Einstein's general relativity is a great success passing through numerous experimental tests. In this century, two predictions of Einstein's gravity, black holes and gravitational waves, were discovered directly~\cite{LIGOScientific:2017vwq,LIGOScientific:2017zic,EventHorizonTelescope:2019dse,EventHorizonTelescope:2019uob,EventHorizonTelescope:2019jan,EventHorizonTelescope:2019ths,EventHorizonTelescope:2019pgp,EventHorizonTelescope:2019ggy}. However, Einstein's gravity may not be the final theory, experimentally, it cannot perfectly explain the acceleration of the universe's expansion, theoretically, it is not renormalizable in the viewpoint of quantum field theory. Many modified gravity theories have been proposed since the foundation of general relativity. A natural extension of  Einstein's gravity theory is adding higher curvature terms. It was shown that  Einstein-Hibbert term plus curvature square terms is renormalizable, though it may suffer from ghost problems~\cite{Stelle:1976gc}.  In order to avoid ghost problems in higher derivative gravity,  chiral gravity~\cite{Li:2008dq}, topological gravity~\cite{Deser:1981wh}, new massive gravity in three dimensions~\cite{Bergshoeff:2009aq} were constructed. Later, Lu and Pope proposed critical gravity in four dimensions, they pointed out that there exists a point in the parameter space of cosmological Einstein's gravity extended with curvature square terms where the theory can avoid ghost problems~\cite{Lu:2011zk}. The critical gravity were then generalized to higher dimensions~\cite{Deser:2011xc}. 

Since the foundation of Einstein's relativity, Einstein realized that it is difficult to construct black hole solutions from the non-linear Einstein's field equations. It is even more difficult to construct black hole solutions in higher derivative gravity theories. However, there is a big breakthrough due to the tremendous advancements in computational power, which can enable us to numerically solve non-linear equations that were previously unsolvable. New static spherical black hole solutions beyond Schwarzschild black hole was constructed in Einstein-Weyl gravity~\cite{Lu:2015cqa}. Then it was generalized to charged case, and two branches of charged black holes were constructed in Einstein-Weyl-Maxwell theory~\cite{Lin:2016jjl,Wu:2019uvq}. Along these successes in constructing black hole solutions are some puzzles which we want to address in this paper: 

(i) What is the relation between the two black hole solutions, Schwarzschild and non-Schwarzschild solutions, in Einstein-Weyl theory and the two branches of charged black hole solutions in Eintein-Weyl-Maxwell theory?

(ii) It is known that Reissner-Nordström (RN)  black hole is not a solution of Einstein-Weyl-Maxwell theory, is there any connection between RN black hole and the two numerically constructed charged black holes?

(iii) Why does the Einstein-Weyl-Maxwell theory not support RN black hole?

We conduct a detailed study about the charged black hole solutions in Einstein-Weyl-Maxwell theory, providing reasonable answers to the first two puzzles and offering an insightful explanation for the third puzzle.

The paper is organized as follows,  we provide a brief review of Einstein-Weyl-Maxwell theory in Section~\ref{section 2}. A detailed analysis of charged black hole solutions is conducted in Section~\ref{section 3}. In Section~\ref{section 4}, we explore the relationship between charged black holes in Einstein-Weyl-Maxwell theory and Reissner-Nordström (RN) black holes. The thermodynamic first law of these two types of charged black holes is carefully examined, and their stability is assessed through the study of quasinormal modes in Section~\ref{section 5}. In Section~\ref{section 6}, we calculate the charge-to-mass ratio of one of the charged black holes in the near extremal limit. We conclude our findings in Section~\ref{section 7}.

\section{Einstein-Weyl-Maxwell theory}\label{section 2}
We consider Einstein gravity extended with quadratic curvature terms plus Maxwell field in four dimensional spacetime, the action is given by
\begin{equation}
	\label{eq2001}
	I=\int \td^4 x \sqrt{-g} \left(R-\alpha C_{\mu \nu \rho \sigma}C^{\mu \nu \rho \sigma} + \beta R^2 - \sigma F_{\mu \nu} F^{\mu \nu}\right),
\end{equation}
where $\alpha,\beta$ and $\sigma$ are coupling constants, $C_{\mu \nu \rho \sigma}$ is Weyl tensor, $F_{\mu \nu} = \nabla_{\mu} A_{\nu}-\nabla_{\nu} A_{\nu}$ is the electromagnetic field strength. The gravity part was studied in~\cite{Lu:2015cqa}, it was pointed out that the theory contains three modes in the linear spectrum, the usual massless spin-2 graviton, a massive spin-2 graviton with mass $m_{2}=1/\sqrt{2\alpha}$, and a massive spin-0 scalar mode with mass $m_{0}=1/\sqrt{6\beta}$. The additional massive modes correspond to Yukawa-type potential terms $\frac{1}{r}\text{e}^{\pm m_{2}r}$ and $\frac{1}{r}\text{e}^{\pm m_{0}r}$ in the metric function in the large $r$ ($r\rightarrow \infty$)\cite{Smilga_2014,Lu:2015cqa,L__2017}.

The equations of motion corresponding to the metric $g_{\mu \nu}$ and gauge field $A_{\mu}$ are given by
\begin{equation}
	\label{eq20012}
	\begin{aligned}
		&R_{\mu \nu}-\frac{1}{2} g_{\mu \nu} R-4 \alpha B_{\mu \nu}+ 2\beta R (R_{\mu \nu}-\frac{1}{4} g_{\mu \nu} R)+ 2\beta (g_{\mu \nu}\square -\nabla_\mu \nabla_\nu )R -8 \pi \sigma T_{\mu \nu} =0, \\
		&\nabla_\mu F^{\mu \nu} =0 .
	\end{aligned}
\end{equation}
where $B_{\mu \nu}$ is the Bach tensor defined as
\begin{equation}
	\label{eq20013}
	B_{\mu \nu}=\left(\nabla^\rho \nabla^\sigma+\frac{1}{2} R^{\rho \sigma}\right) C_{\mu \rho \nu \sigma},
\end{equation}
and $T_{\mu \nu}$ is the energy-momentum tensor of the Maxwell field
\begin{equation}
	\label{eq20014}
	T_{\mu \nu}=F_{\mu \alpha} F_\nu{ }^\alpha-\frac{1}{4} g_{\mu \nu} F_{\rho \sigma} F^{\rho \sigma} .
\end{equation}
Taking the trace of the Einstein equations in \eqref{eq20012} gives
\begin{equation}
	\label{eq20015}
	\left(\square-m_0^2\right)R=0 .
\end{equation}
One can notice that the Maxwell field doesn't contribute to the trace part equation, that is because the trace of the energy momentum tensor of Maxwell field is zero in four dimension. 
In fact, Equation \eqref{eq20015} is the equation of the spin-0 scalar mode. It was pointed out that the massive scalar mode is forbidden for a static and spherical black hole. The reason can be summarised as follows. Considering a static, spherically symmetric metric $\td s_{4}^{2}=-h(r)  \td t ^{2}+\td r^{2}/f(r)+r^{2} (\td \theta^2 +\sin^2\theta \td \phi^2)$. Multiplying the equation\eqref{eq20015} by Ricci scalar $R$ and integrating it from the event horizon to spatial infinity yields 
\begin{equation}
	\label{eq20016}
	\begin{aligned}
	&\int \td^3 x \sqrt{-g} R\left(\square-m_{0}^{2}\right)R\\
	 = &\int \td^3 x \left[\partial_{r}\left(\sqrt{-g} g^{rr}R \partial_{r} R \right)-\sqrt{-g} \left(g^{rr}\partial_{r} R\partial_{r} R + m_{0}^2 R^2\right)\right]=0 .
	\end{aligned}
\end{equation}

The first term of this integral is a total derivative, while the last two terms are quadratic terms. Since $\sqrt{-g} g^{rr}R \partial_{r} R$ vanishes at both the event horizon and spatial infinity, the contribution from the total derivative term disappears. And $\sqrt{-g}$ and $g^{rr}$ are always positive outside the event horizon. Therefore, it implies that both the second and third terms must vanish, leading to $R=0$. In other words, the existence of black hole solutions requires the vanishing of the Ricci scalar. As is mentioned, the Maxwell field don't contribute to the trace part, thus this property of vanishing Ricci scalar still holds for the theory~\eqref{eq2001}. Thus we shall not consider the $R^2$-term by setting $\beta=0$.                                                                                                                                                                                              

And  without loss of generality, we set $\sigma=1$, the action \eqref{eq2001} turns out to be 
\begin{equation}
	\label{eq2002}
     I=\int \td^4 x \sqrt{-g} \left(R-\alpha C_{\mu \nu \rho \sigma}C^{\mu \nu \rho \sigma}  - F_{\mu \nu} F^{\mu \nu}\right),
\end{equation}
which is called Einstein-Weyl-Maxwell theroy.

We aim to explore the static spherical black holes with in this theory, with the ansatz
\begin{equation}
	\label{eq2005}
	\begin{aligned}
	\td s_{4}^{2}&=-h(r)  \td t ^{2}+\frac{\td r^{2}}{f(r)}+r^{2} (\td \theta^2 +\sin^2\theta \td \phi^2), \\
		A &= - a (r)\td t .
	\end{aligned}
\end{equation}
The metric $h,f$ and electric potential $a$ are functions of only radial coordinate $r$.  Through the Maxwell equations, the electric potential $a$ can be connected to the metric functions
\begin{equation}
	a'=\sqrt{\frac{h}{f}}\frac{Q}{r^2}.
\end{equation}	
Hereafter, a prime "$'$" denotes derivative with respect to radial coordinate $r$. $Q$ is an integration constant, which can be interpreted as the electric charge of the black hole.

Together with condition $R=0$ , the Einstein equations in \eqref{eq20012} turns to 
\begin{equation}
	\label{eq2006}
	\begin{aligned}
		h''&+\frac{h'}{2}\left(\frac{f'}{f}-\frac{h'}{h}\right)+\frac{2}{r f}\left(h f\right)'+\frac{2h}{r^2 f}\left(f-1\right)=0,\\
		f''&-\frac{r\left(h'\right)^2}{2\left(r h'-2h\right)}\left(\frac{f}{h}\right)'+\frac{3 h^2\left(f'\right)^2+2 h f h' f'+3 f^2\left(h'\right)^2}{2h f\left(r h'-2h\right)}\\
		&+\frac{2\alpha \left(f-1\right)h f'-r^2 f h'}{\alpha r f\left(r h'-2h\right)}+\frac{h \left[\left(4\alpha f-r^2\right)\left(f-1\right)-Q^2\right]}{\alpha r^2 f\left(r h'-2h\right)}=0
	\end{aligned}
\end{equation}

As is  discussed in Reference~\cite{Lu:2015cqa}, the famous Schwarzschild black hole is a solution to the Einstein-Weyl theory. Here, for Einstein-Weyl-Maxwell theory, the well-known Reissner-Nordström (R-N) black hole is not a solution anymore. And as far as we known there are no analytical static black hole solutions to the Einstein-Weyl-Maxwell theory. 

Though, analytical black hole solutions are difficult to construct in Einstein-Weyl-Maxwell theory, one can obtain the asymmptotic behavior of the black hole, if there does exist black hole solutions. In the large $r$ region, the black hole is supposed to be asymmptotic to Minkovski spacetime and have the form of 
\begin{equation}
	\label{eq20062}
	\begin{aligned}
		&h(r)=h_0\left(1+\delta h(r)+\mathcal{O}\left(\delta h(r)^2\right)\right),\\
		&f(r)=1+\delta f(r)+\mathcal{O}\left(\delta f(r)^2\right),\\
		&a(r)=\delta a(r)+\mathcal{O}\left(\delta a(r)^2\right).\\
	\end{aligned}
\end{equation}
Here,  $h_{0}$ is a finite constant which can be set to 1 through the scaling symmetry of time coordinate. Solving the equations of motion to linear order, one can obtain
\begin{equation}
	\label{eq20063}
	\begin{aligned}
		&h(r)=h_0\left[1-\frac{C_{0}}{r}-C_{1}\frac{\text{e}^{-\frac{r}{\sqrt{2\alpha}}}}{r}+\frac{Q^2}{r^2}\right]+\mathcal{O}\left(\frac{1}{r^4}\right),\\
		&f(r)=1-\frac{C_{0}}{r}-C_{1}\left(1+\frac{\sqrt{2\alpha}}{r}\right)\text{e}^{-\frac{r}{\sqrt{2\alpha}}}+\frac{Q^2}{r^2}+\mathcal{O}\left(\frac{1}{r^4}\right),\\
		&a(r)=-\sqrt{h_{0}}\frac{Q}{r}+\mathcal{O}\left(\frac{1}{r^5}\right).\\
	\end{aligned}
\end{equation}
Where, $C_{0}$ is an integration constant associated with the mass and $C_{1}$ is an integration constant corresponding to the massive spin-2 modes, and we have get rid of the divergent mode $\text{e}^{+\frac{r}{\sqrt{2\alpha}}}$ by setting its coefficient to zero.

At the event horizon $r_0$, we have $h(r_0)=f(r_0)=0$. We can perform a Taylor expansion of $h(r)$ and $f(r)$ near the horizon,
\begin{equation}
	\label{eq2007}
	\begin{aligned}
		&h(r)=h_1(r-r_{0})+h_2(r-r_{0})^2+h_3(r-r_{0})^3+...,\\
		&f(r)=f_1(r-r_{0})+f_2(r-r_{0})^2+f_3(r-r_{0})^3+...
	\end{aligned}
\end{equation}
Through the  field equations \eqref{eq2006}, higher-order coefficients $\{f_i,\, h_i\}$ with $i\ge 2$,  can be expressed in terms of $\{h_1,\,f_1, \, r_0\}$. The second-order coefficients are given by
\begin{equation}
	\label{eq2008}
	\begin{aligned}
		h_2=&\frac{1}{8 \alpha f_1^2 r_{0}^3} \left(-Q^2 f_{1} +h_{1} r_{0} \left(-16 \alpha  f_{1}^2 r_{0}+8 \alpha  f_{1}-f_{1} r_{0}^2+r_{0}\right)\right),\\
		f_2=&\frac{1}{8 \alpha h_1 f_1 r_{0}^3}\left( 3 Q^2 f_{1} +h_{1} r_{0} \left(-r_{0} \left(16 \alpha  f_{1}^2+3\right)+8 \alpha  f_{1}+3 f_{1} r_{0}^2\right) \right).
	\end{aligned}
\end{equation}
Here,again, the coefficient $h_1$ can be set to $r$ through scaling symmetry.  Then there are total three integration constants $\{ f_1,\, r_0\,,Q\}$ corresponding to three modes, the usual massless spin-2 graviton, the massive spin-2 gravity mode and the spin-1 vector mode.

\section{Classification of Black Holes in Einstein-Weyl-Maxwell theory}\label{section 3}

\begin{figure}[H]	
	\centering
	\includegraphics[width=0.45\textwidth]{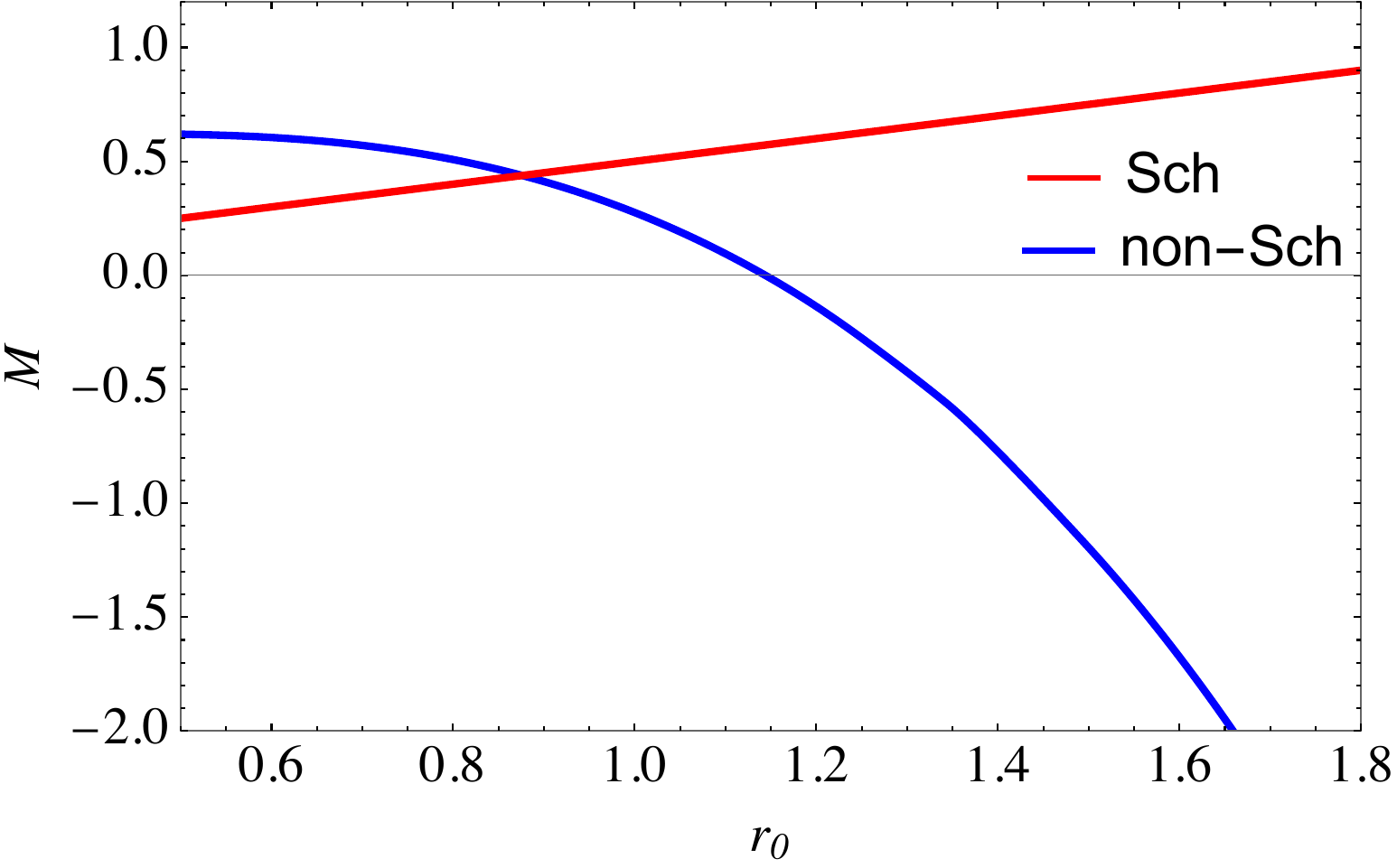}
	\qquad
	\includegraphics[width=0.45\textwidth]{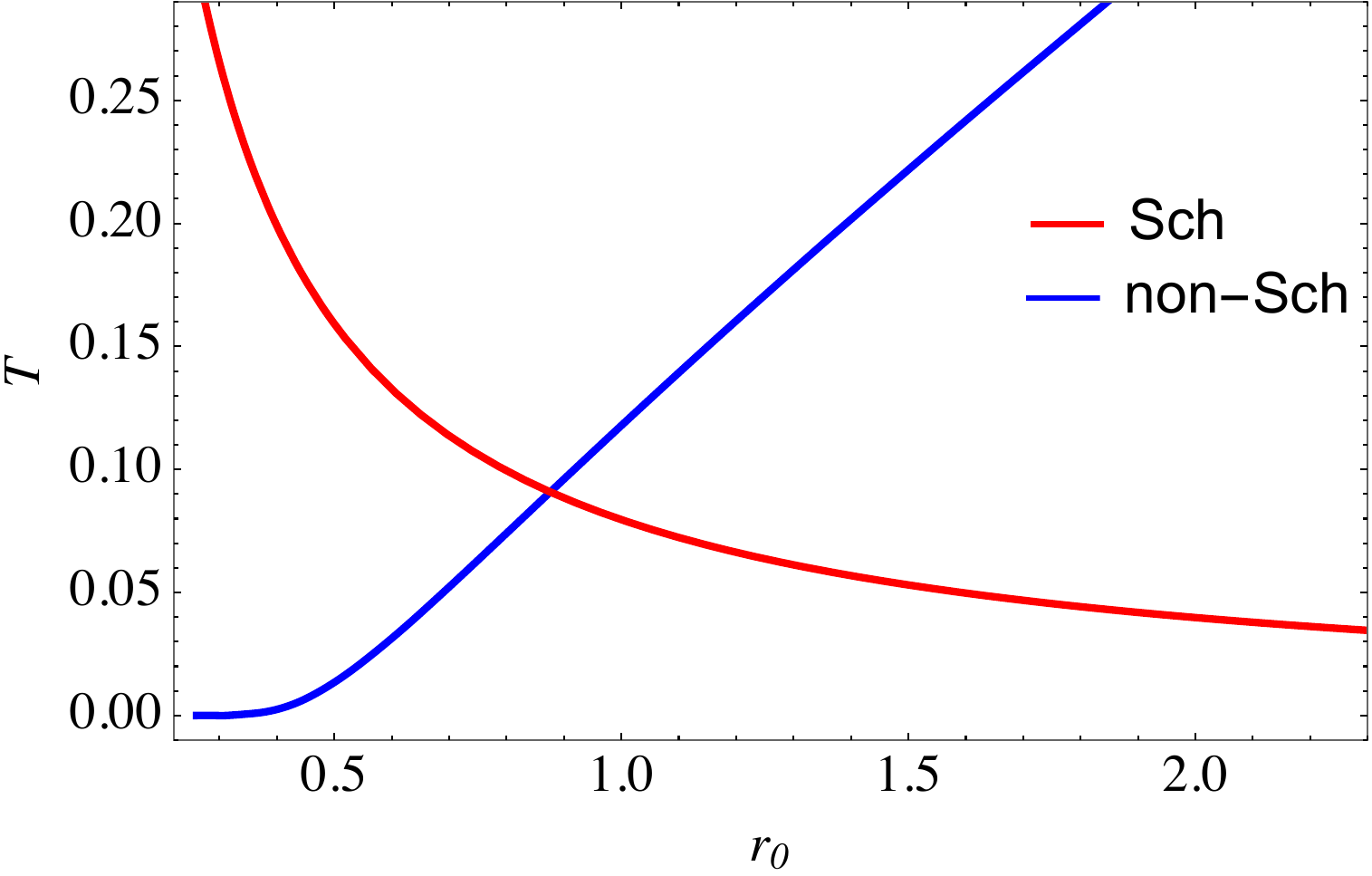}
	\caption{The mass and Temperature (M, T) as  functions of event horizon radius $r_0$ in shown. There are two branches of solutions in Einstein-Weyl theory, one is Schwarzschild black hole the other is called non-Schwarzschild black hole. The two branches of solutions intersect with each other.  }
	\label{Fig3000101}
\end{figure}

Constructing analytic black holes in higher derivative gravity theories is tough, then people turn to numerical methods. It was pointed out that a new static black hole  was found through numeric tools in Einstein-Weyl theory~\cite{Lu:2015cqa}. Since Schwarzschild black hole is automatically a solution in Einstein-Weyl theory, thus there are two branches of solutions in Einstein-Weyl theory, as is shown in Figure~\ref{Fig3000101}. The new solution was called non-Sch in Figure~\ref{Fig3000101}. It obvious that the two branches intersect with each other.

Later, people went a step further to consider Einstein-Weyl-Maxwell theory and also constructed two groups of charged black holes. They thought one of them corresponds to the Schwarzschild black hole and the other is a generalization of non-Schwarzchild black hole in the Einstein-Weyl gravity~\cite{Lu:2015cqa}. However, the connection between the charged black hole and uncharged one is not clear. We carried a detailed study of the charged black holes and give a natural classification of these two branches of charged black holes. Then the connection between the charged and uncharged black hole was shown more intuitive.

First, we want to present the charged black holes in Einstein-Weyl-Maxwell theory. We choose to integrate the field equation from horizon to infinity, since the Taylor expansion~\eqref{eq2007} can be carried out to a very high order which can be used as the initial conditions for numerical shooting method.

\begin{figure}[H]	
	\centering
	\includegraphics[width=0.45\textwidth]{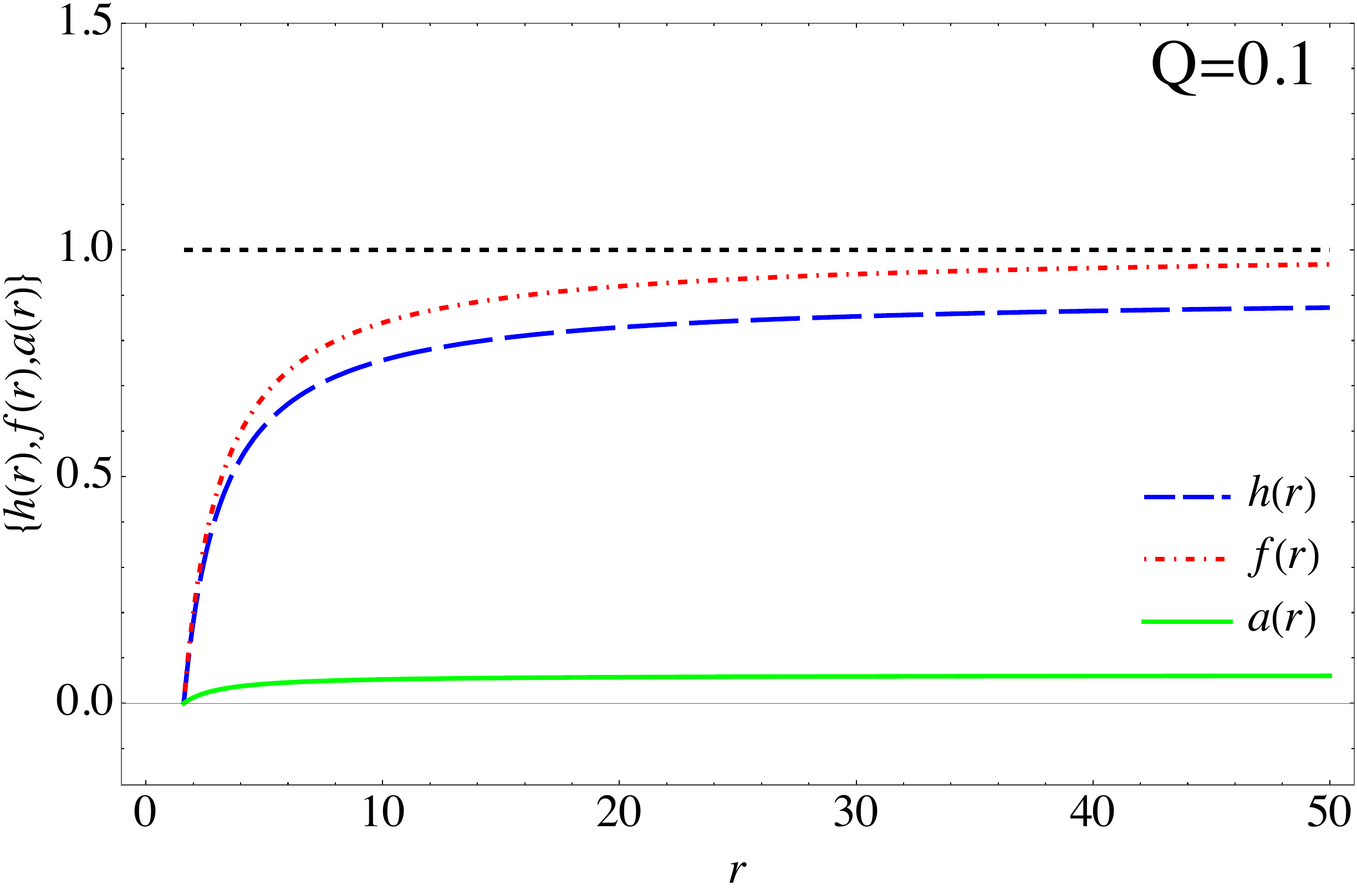}
	\qquad
	\includegraphics[width=0.45\textwidth]{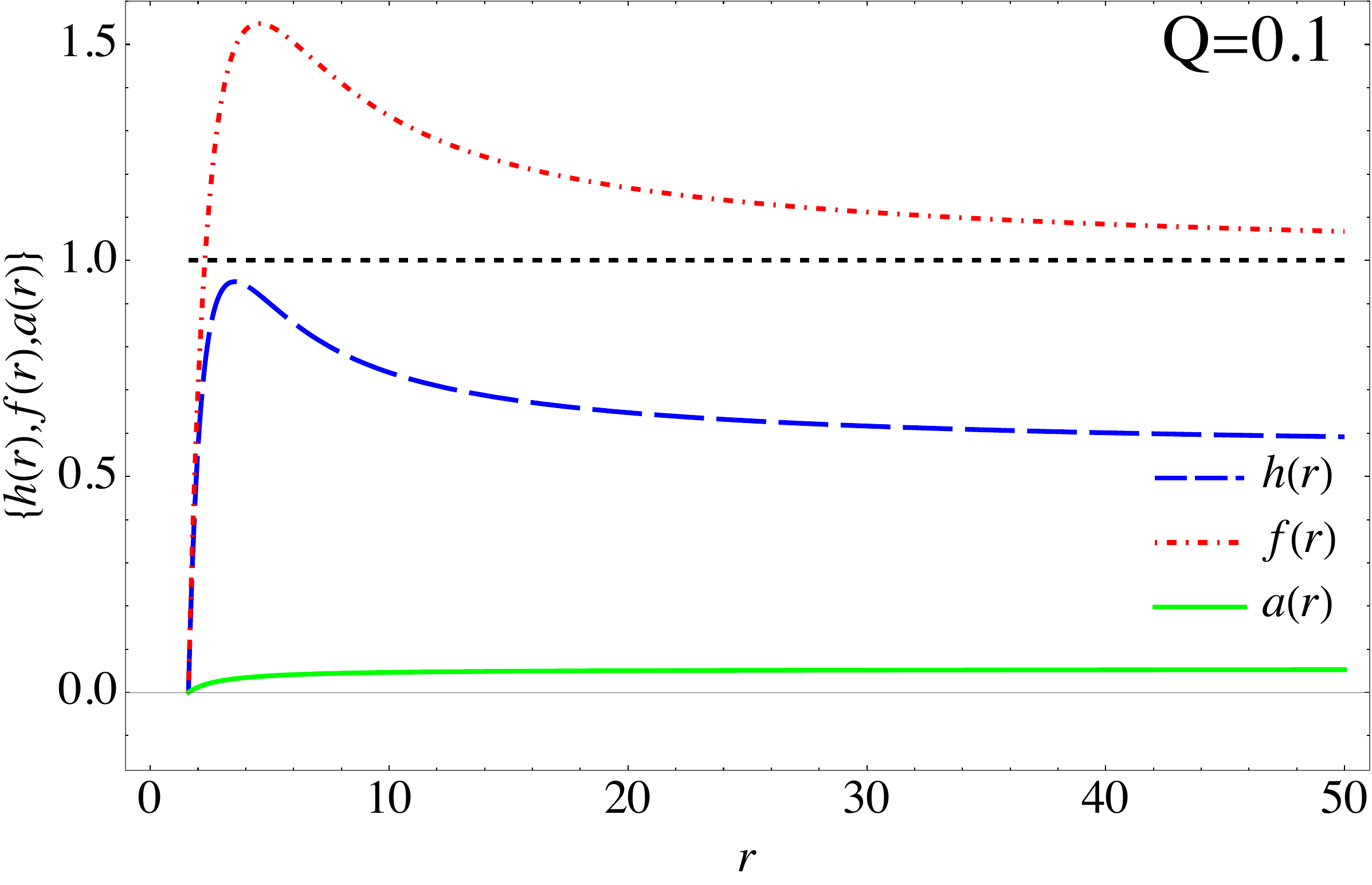}
	\caption{The metric functions $h(r), f(r)$ and the electric potential $a(r)$ as functions of $r$, with electric charges $Q=0.1$ and horizon radius $r_{0}=1.6$. The profile of the metric functions has no peak in the left panel, whilst there is a peak in the right panel.  }
	\label{Fig30001}
\end{figure}

We make the same choice of the theory parameter $\alpha = 1/2$ as that of uncharged case~\cite{Lu:2015cqa} and choose a small charge $Q=0.1$ as the first example. The profile of the metric functions and electric potential $\{ h(r),\, f(r),\, a(r)\}$  are shown in Figure~\ref{Fig30001}. As is can be seen from Figure~\ref{Fig30001}, the profile of metric functions have no peak in the left panel, for which the black hole has a positive mass. Whist there is a peak in the metric profile in the right panel, whose mass is negative. Since the black hole we studied is asymmptotic to Minkovshi spacetime, the mass of the black hole can be read off from the fall of  $g_{tt} = - (1 - 2 m/r+ \dots )$.

\begin{figure}[H]	
	\centering
	\includegraphics[width=0.75\textwidth]{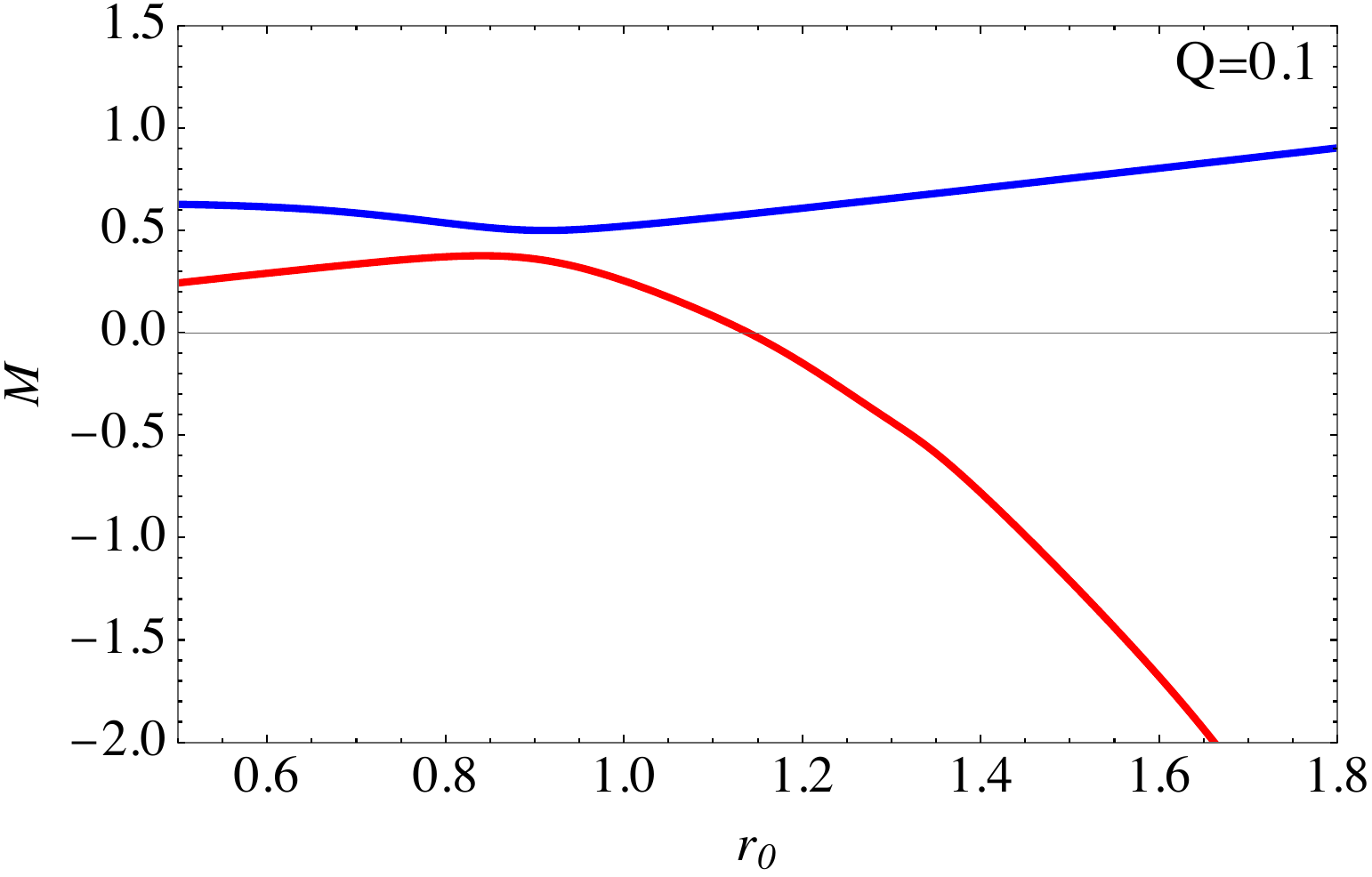}
	\caption{The mass as a function of event horizon radius for the two branches of black hole solutions, with electric charges $Q=0.1$. The two branches don't intersect with each other. }
	\label{Fig30021}
\end{figure}

These two black holes have the same electric charge $Q=0.1$ and the same event horizon radius $r_0 = 1.6$, but have different mass. This implies there exist two branched of solution in the Einstein-Weyl-Maxwell theory. Accumulating more data, the $M-r_0$ curves for the two branches of black hole solutions are displayed in Figure~\ref{Fig30021}.

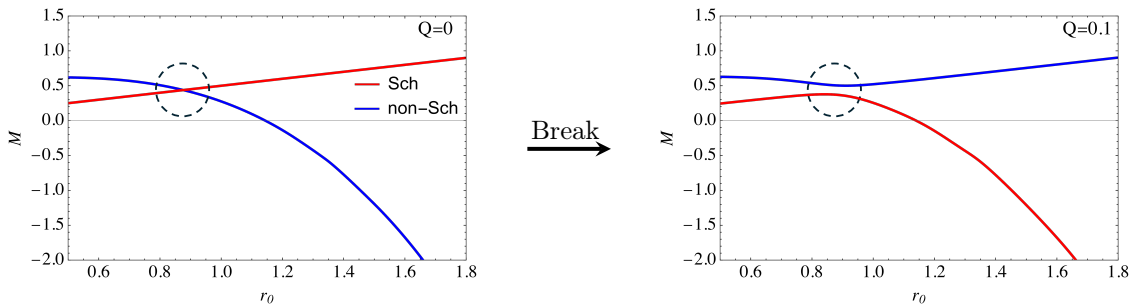
\begin{figure}[H]
	\centering
	% 第一张图片
	\raisebox{-0.5\height}{\includegraphics[width=0.42\textwidth]{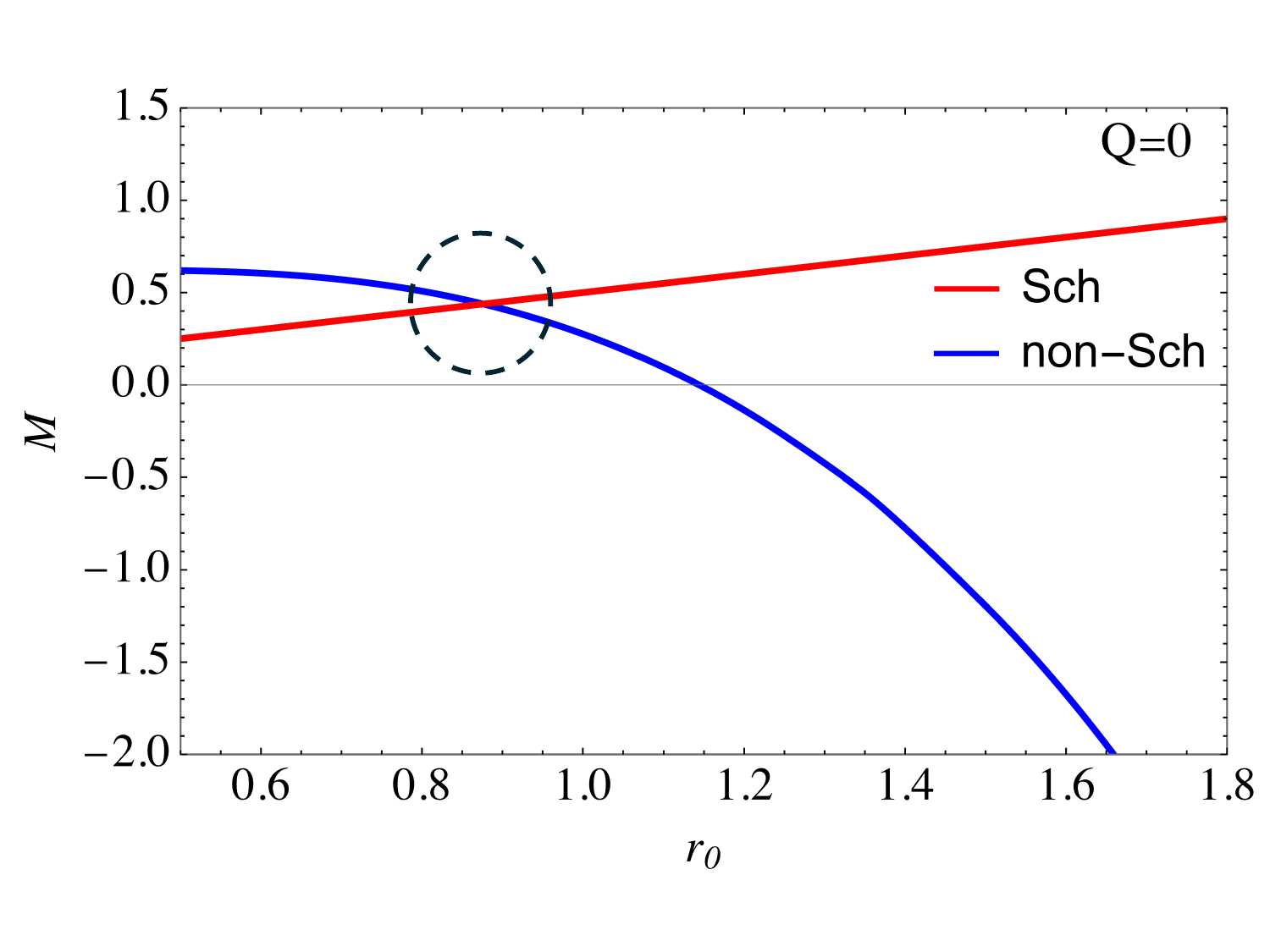}}
	% TikZ 箭头（调整 y 坐标）
	\quad
	\begin{tikzpicture}[baseline=-0.5ex]
		\draw[-stealth, line width=1.5pt] (0,0) -- (1.1,0);
		\node[above, font=\small] at (0.5,0) {Break};
	\end{tikzpicture}
	\quad
	% 第二张图片
	\raisebox{-0.5\height}{\includegraphics[width=0.42\textwidth]{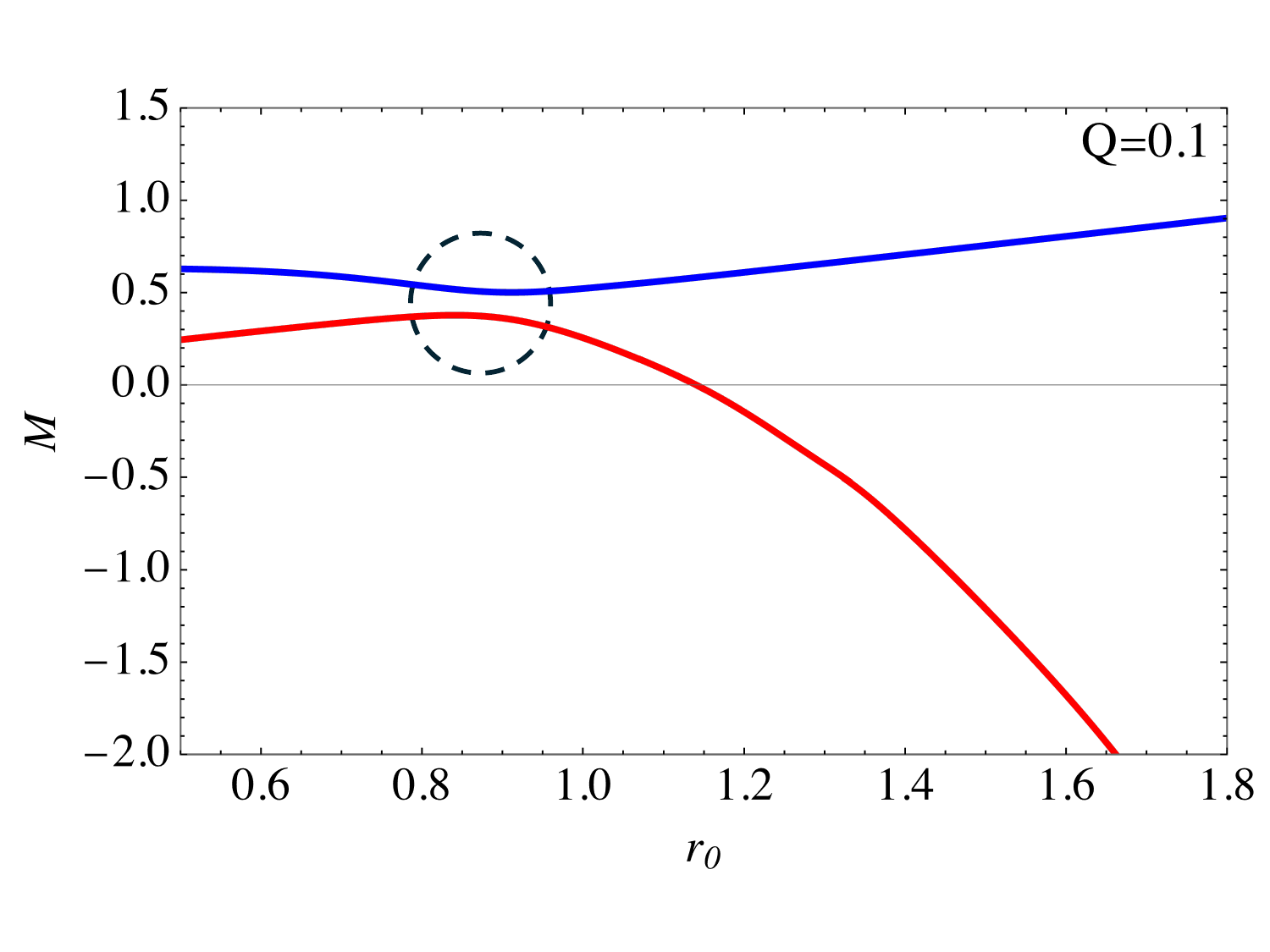}}
	\caption{ There is a big difference between the charged case and the neutral one, the branches of neutral solutions intersect with each other whilst the two branches of charged solutions don't intersect with each other. Adding charge triggers the the neutral ones to break off from the intersection to two disconnected branches of charged solution. }
	\label{Fig30041}
\end{figure}

Before our work, people thought that the charged case is a simple generalization of the uncharged case, explicitly,  one branch of the solution is a generalization of Schwarzschild black hole, the other is related to non-Schwarzschild black hole~\cite{Lu:2015cqa}. However, we find out that  this kind of classification is inappropriate. In contrast, as is can be seen from Figure~\ref{Fig30021}, each branches of charged black hole shares half of Schwarzschild and half of non-schwarzschild black hole.  And the procedure can be manifested through Figure~\ref{Fig30041}. When there is no charge, the Schwarzschild solution and the non-Schwarzschild solution intersect at one point.  Whilst turning on the charge, the two branches of neutral solutions break off from the intersection point to two disconnected branches of charged black holes, see Figure~\ref{Fig30041}. 

Next, we turn to temperature and entropy of the charged black holes. The temperature of the black hole can be calculated through the standard procedure
\begin{equation}
	\label{eq2015}
	T=\frac{\sqrt{h^{\prime}(r_{0})f^{\prime}(r_{0})}}{4\pi}=\frac{\sqrt{h_{1}f_{1}}}{4\pi}.
\end{equation}
The entropy of the black hole can be obtained by the Wald entropy formula \cite{Wald1983,IyerWald1994}. For static spherical black holes in Einstein-Weyl-Maxwell theory, the entropy has a simple form 
\begin{equation}
	\label{eq20161}
	S=\pi r_{0}^2\left(1-\frac{\alpha}{3}\left(\frac{4(1+r_{0}f^{\prime}(r_{0}))}{r_{0}^2+n^2}-\frac{3f^{\prime}(r_{0})}{2h^{\prime}(r_{0})}h^{\prime\prime}(r_{0})-\frac{1}{2}f^{\prime\prime}(r_{0})\right)\right).
\end{equation}
The Maxwell field has no direct contribution to the entropy. By using the relations of  coefficients in \eqref{eq2008}, the entropy of the black hole can be further simplified
\begin{equation}
	\label{eq2016}
	S=\pi r_{0}^2-4\pi \alpha r_{0}f_{1}.
\end{equation}

It is well-known that the Gauss-Bonnet combination is a total derivative in four dimension, thus adding a Gauss-Bonnet term to Einstein-Weyl-Maxwell theory \eqref{eq2002} will not change the equations of motion, but the Gauss-Bonnet term could have a constant contribution to the entropy. Thus, we can use this freedom to make the entropy above reduce to the entropy of Schwarzschild black hole when $\alpha =0$.

\begin{figure}[H]	
	\centering
	\includegraphics[width=0.45\textwidth]{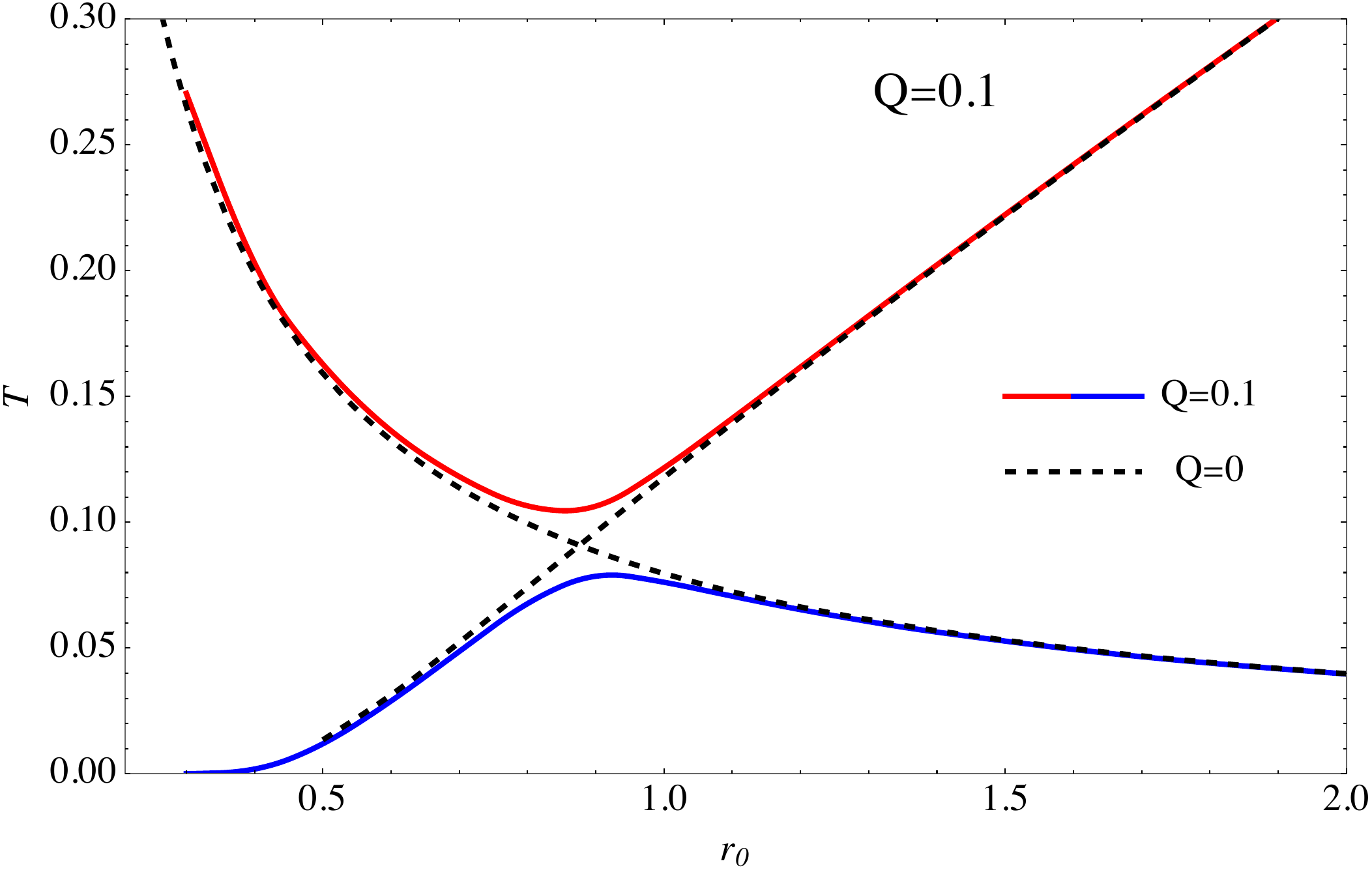}
	\qquad
	\includegraphics[width=0.45\textwidth]{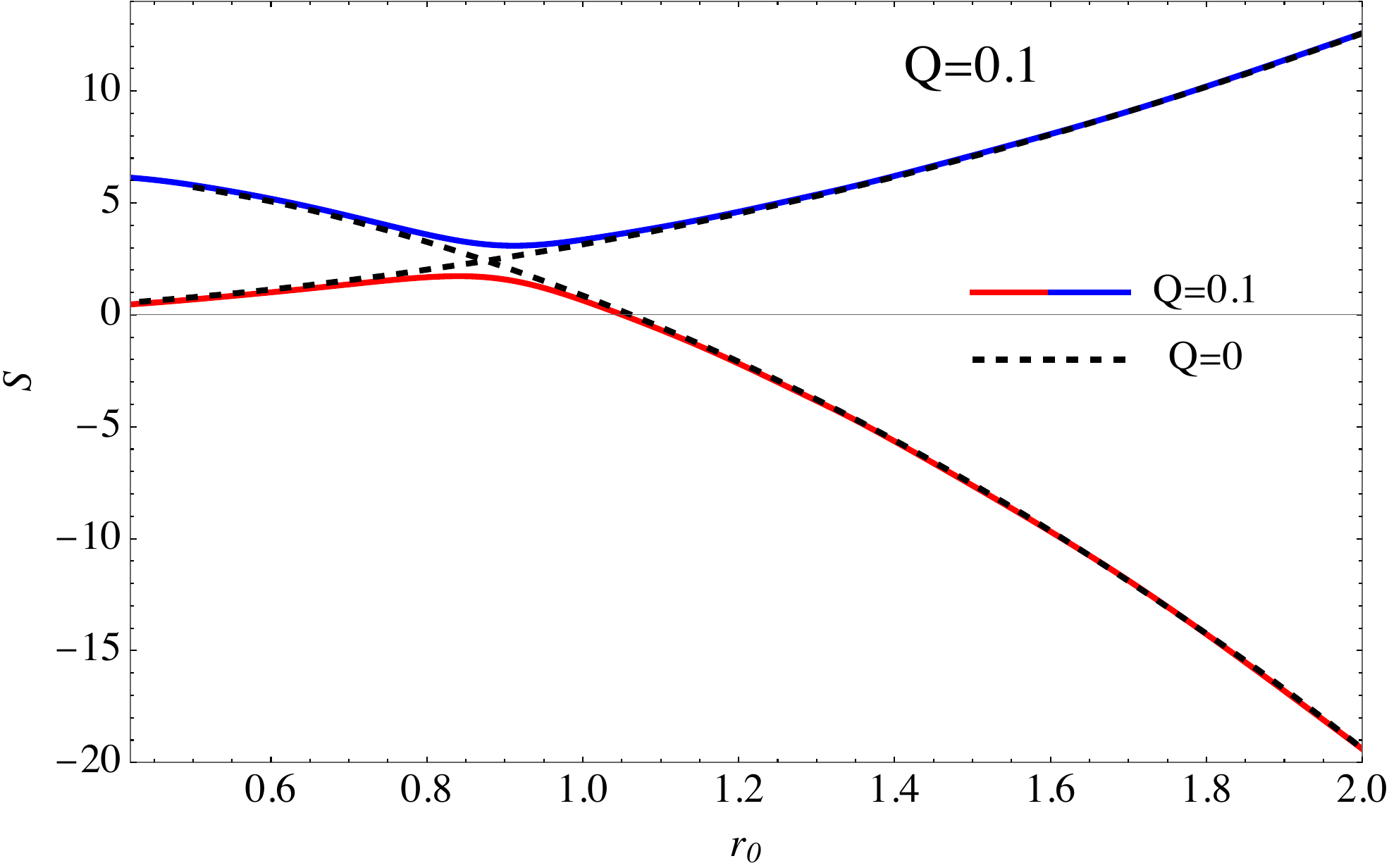}
	\caption{The temperature and the entropy  as functions of horizon radius for fixed charge $Q=0.1$ are shown in the left and right panel respectfully. The temperature and entropy of  the two neutral solutions (dashed line) are also plotted as a comparison. }
	\label{Fig3005}
	
\end{figure}

The temperature and entropy as functions of the event horizon radius for fixed charge $Q=0.1$ are illustrated in the Figure~\ref{Fig3005}. As a comparison,  the uncharged case $Q=0$,  the Schwarzschild and non-Schwarzschild solution, was also presented. From the plots of temperature and entropy, it can  also be observed that the presence of charge $Q$ causes the neutral Schwarzschild and non-Schwarzschild solutions to break from the intersect point into two new numerical solutions. Unlike the neutral case, these two solutions do not intersect.

\begin{figure}[H]	
	\centering
	\includegraphics[width=0.75\textwidth]{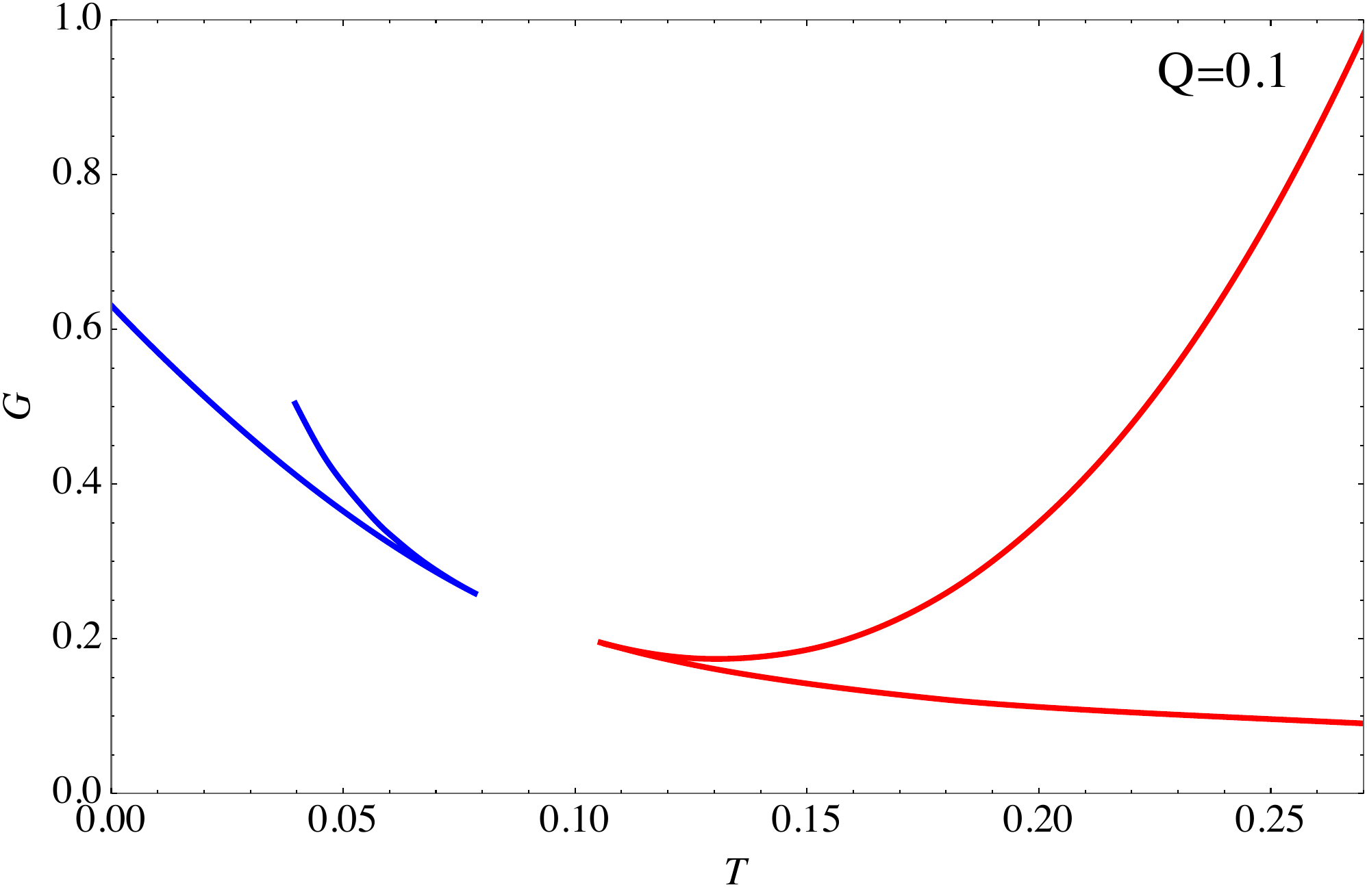}
	\caption{The free energy $G$ as a function of temperature $T$ for fixed charge  $Q=0.1$. The two branches of solutions are disconnected. }
	\label{Fig30022}
\end{figure}

And the free energy $G$ as a function of temperature is depicted in Figure~\ref{Fig30022}, it is obvious that the two branches of solutions are disconnected.

After a careful study of the two branches of charged black hole solutions in Einstein-Weyl-Maxwell theory, we first discovered that these two branches of black solutions  don't intersect, the pattern is totally different from that of the uncharged case. Thus, we point out that the classification of the two branches solutions correspond to the Schwarzchild solution and non-Schwarzchild solution is inaccurate. We unveil the relations between the charged solutions and uncharged one. The charge causes the intersected uncharged solutions to break from the intersection point into two charged solutions, which shows the charge has a very important role in this procedure. 

We shall present two more examples to confirm the pattern we discovered, one is smaller charge $Q=0.03$ and the other is larger charge $Q=1$. 

\begin{figure}[H]	
	\centering
	\includegraphics[width=0.45\textwidth]{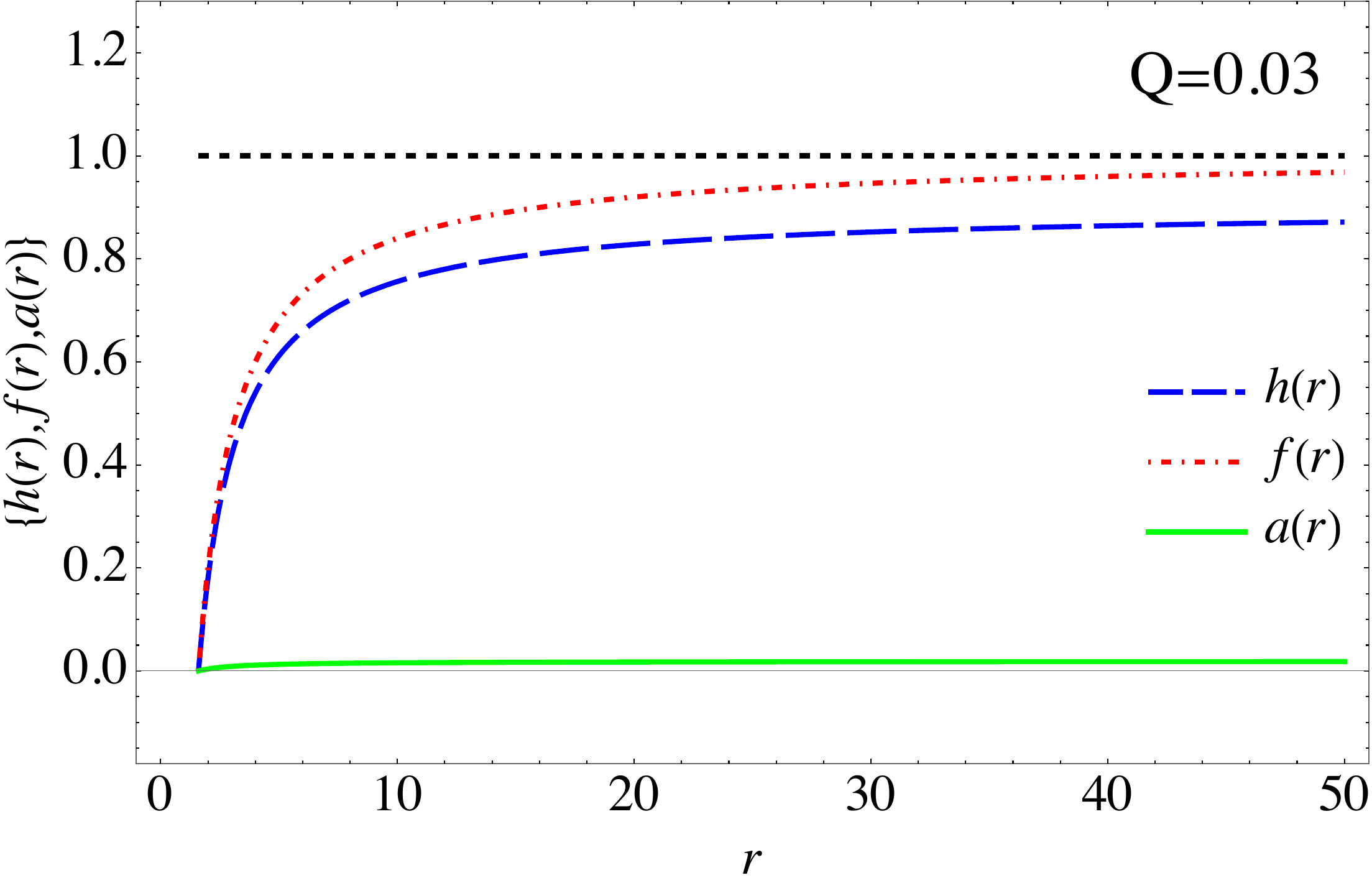}
	\includegraphics[width=0.45\textwidth]{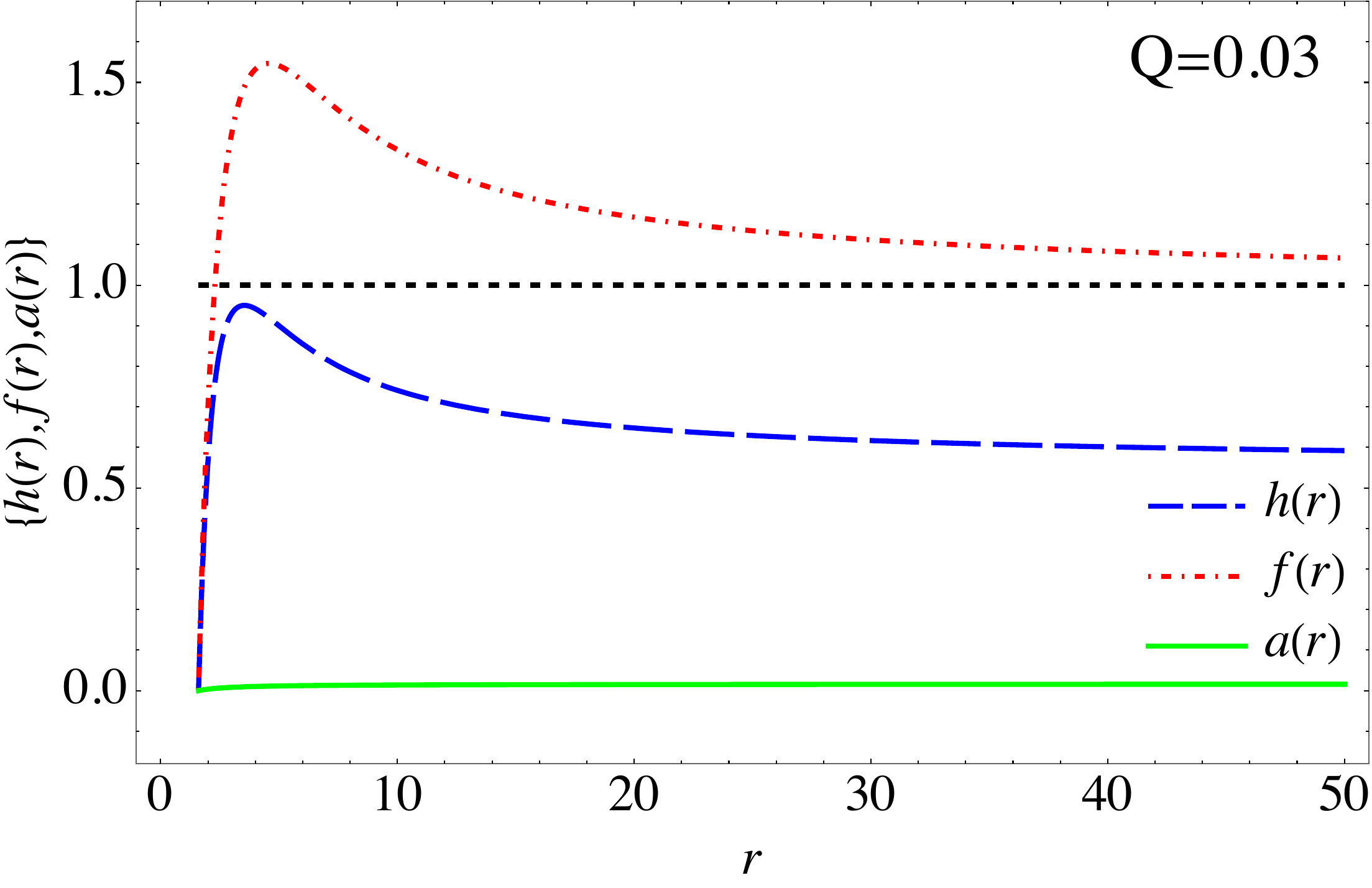}
	\caption{  The metric functions $h(r), f(r)$ and the electric potential $a(r)$ as functions of $r$, with electric charges $Q=0.03$. The profile of the metric functions has no peak in the left panel, whilst there is a peak in the right panel.  }
	\label{Fig30061}
\end{figure}

For $Q=0.03$, the profile of metric function with negative and positive mass was displayed in Figure~\ref{Fig30061}.

\begin{figure}[H]	
	\centering
	\includegraphics[width=0.75\textwidth]{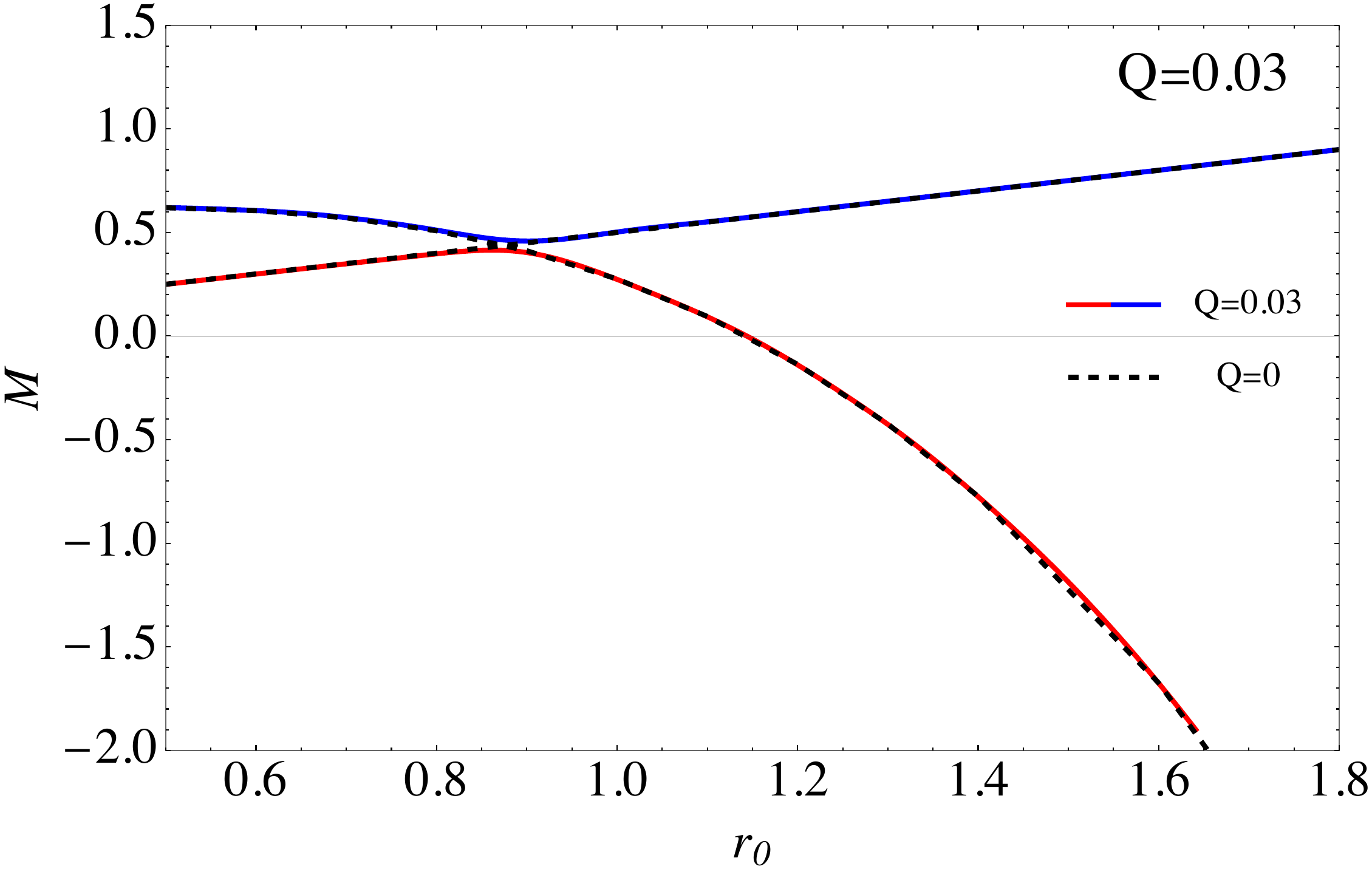}
	\caption{The figure shows  mass as a function of horizon radius for fixed  electric charges $Q=0.03$, and the case for neutral solutions is also plotted as a comparison.}
	\label{Fig3007}
\end{figure}

The mass as a function of horizon radius $r_0$ is plotted in Figure~\ref{Fig3007}, where the uncharged Schwarzschild and non-Schwarzschild were also included. The pattern about charged black holes and neutral ones is the same. Due to the smallness of the charge, the charged black hole solutions almost coincide with the neutral ones, however the two branches of charged black hole don't cross over each other, they break up in the crossover of the uncharged black holes. 

\begin{figure}[H]	
	\centering
	\includegraphics[width=0.45\textwidth]{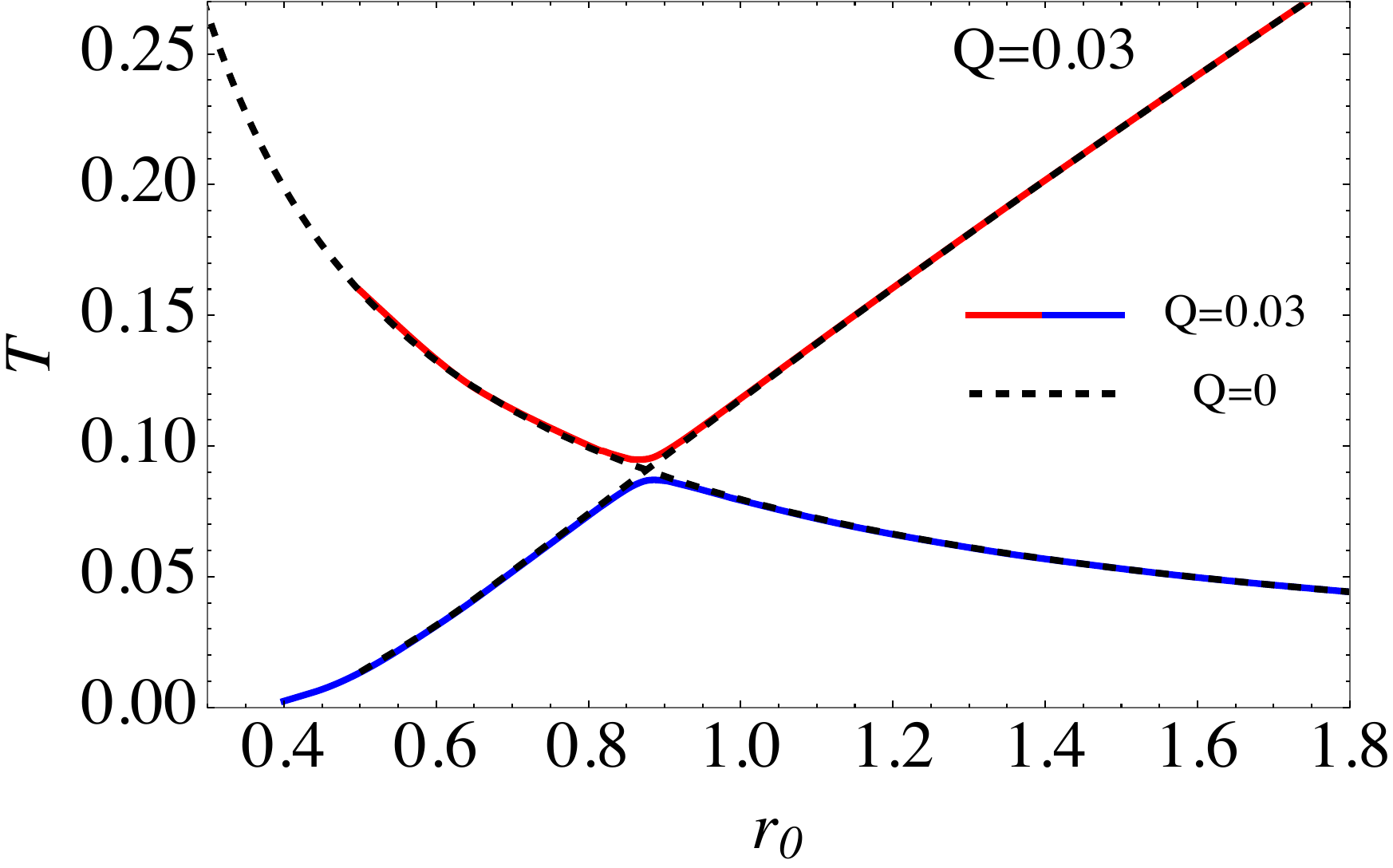}
	\includegraphics[width=0.45\textwidth]{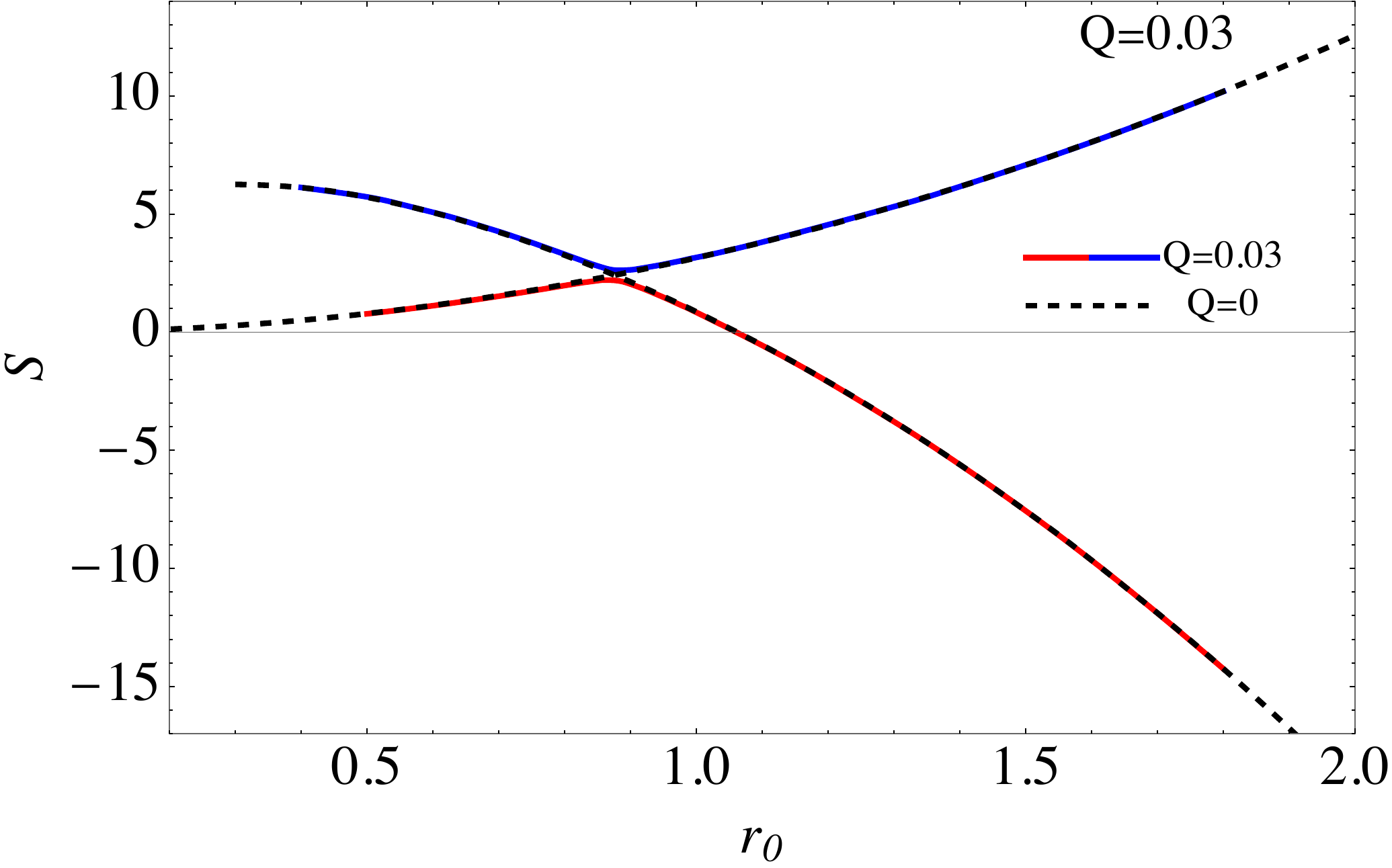}
	\caption{ The temperature and entropy are plotted as functions of horizon radius for fixed charge $Q=0.03$, and the case for neutral solutions are included, too. The charged and uncharged solutions almost overlap with each other. }
	\label{Fig3006}
\end{figure}

The temperature and entropy as functions of the event horizon radius are illustrated in the Figure~\ref{Fig3006}. The dashed lines represent the uncharged black hole solutions, it is within expectation that the two branches of the charged solutions don't intersect.  

\begin{figure}[H]	
	\centering
	\includegraphics[width=0.75\textwidth]{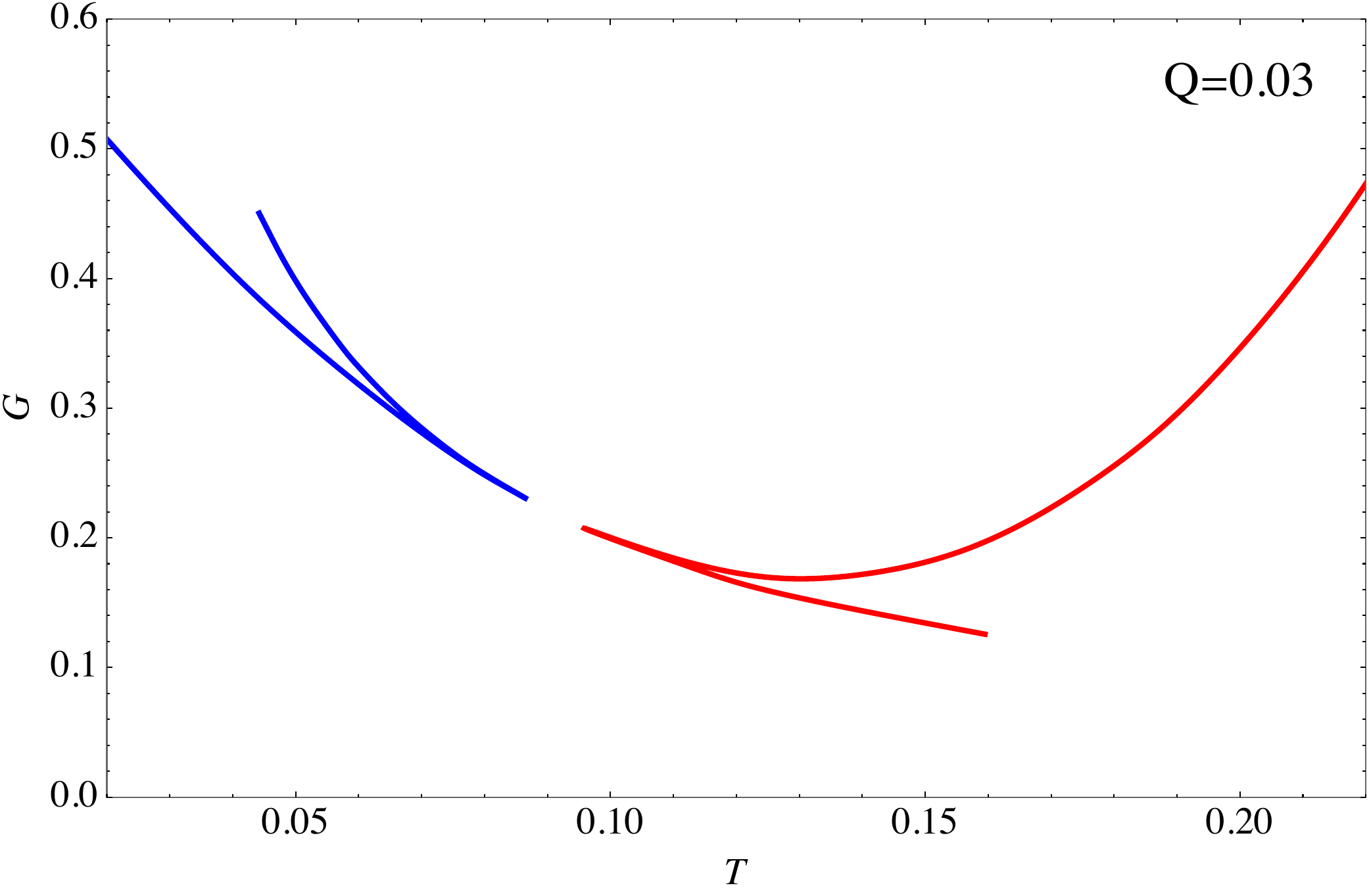}
	\caption{The figure shows the free energy $G$   as a function of temperature $T$ for $Q=0.03$, the two branches of charged solutions are close but disconnected.} 
	\label{Fig30072}
\end{figure}

 And the relationshp of temperature $T$ and free energy $G$ is depicted in the Figure~\ref{Fig30072}, which shows the two branches of solutions are disconnected. 

\begin{figure}[H]	
	\centering
	\includegraphics[width=0.43\textwidth]{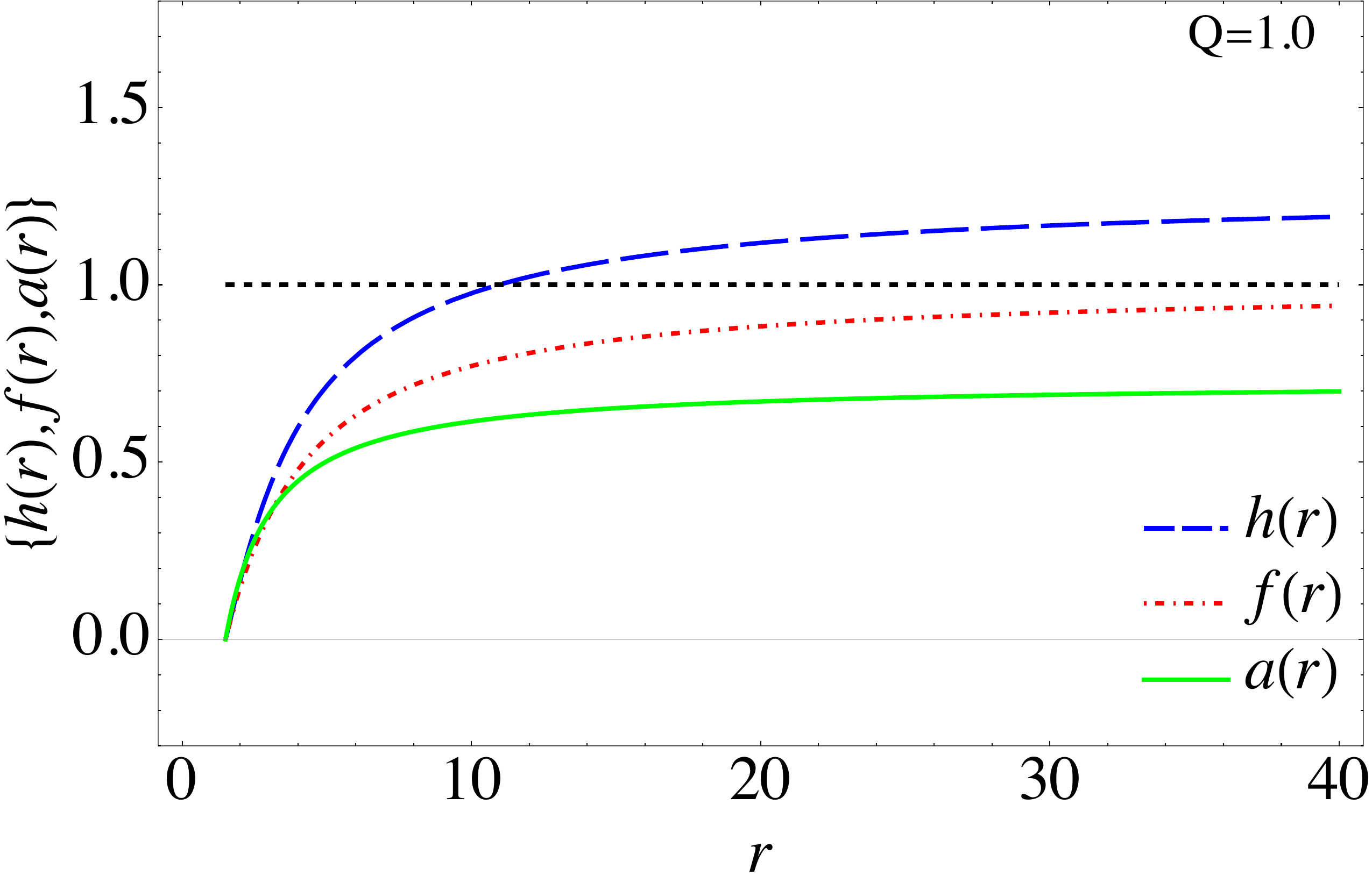}
	\qquad
	\includegraphics[width=0.43\textwidth]{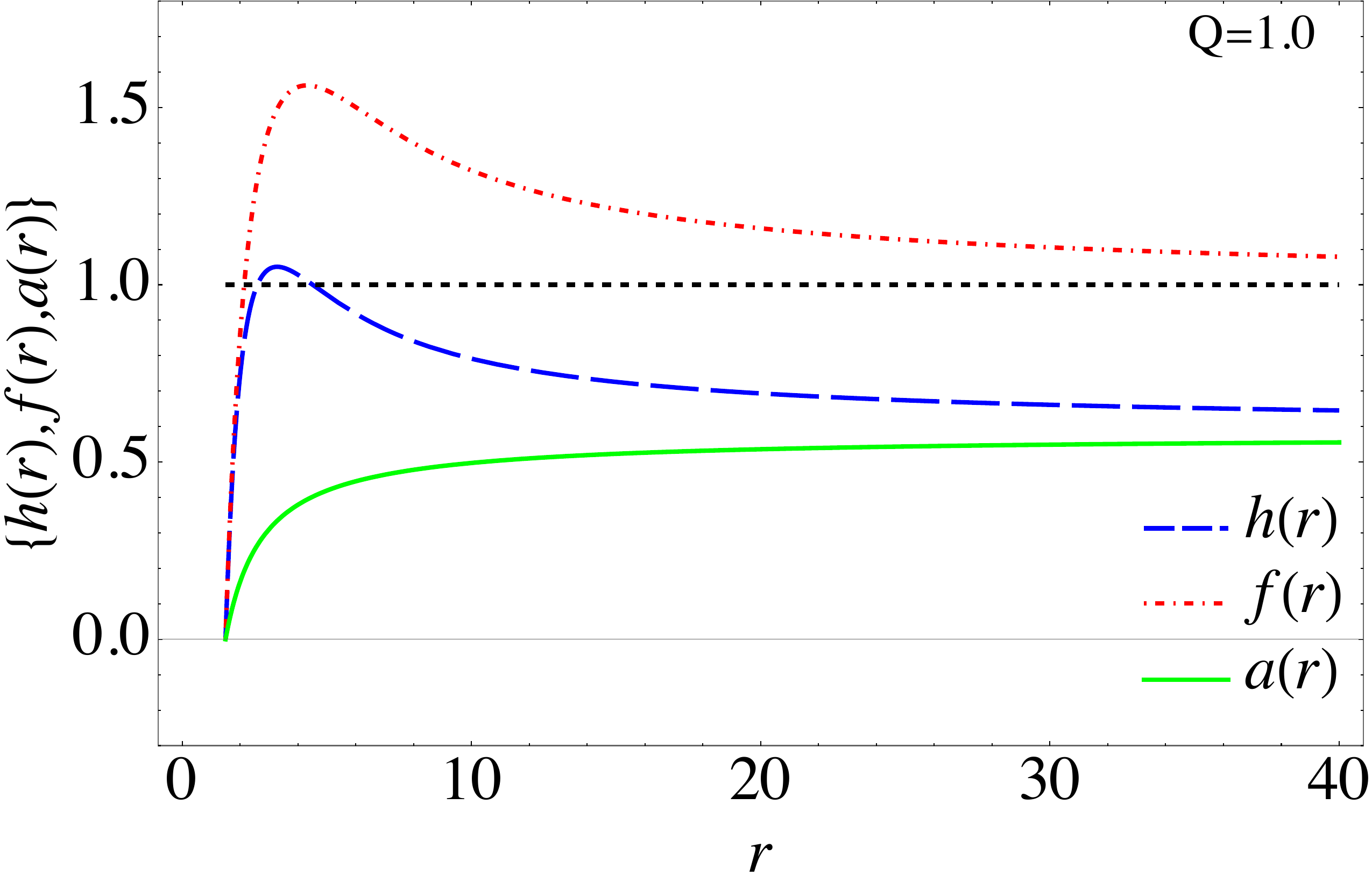}
	\caption{The metric functions $h(r), f(r)$ and electric potential $a(r)$ as functions of $r$ are plotted for $Q=1.0$ and horizon radius $r_{0}=1.5$. The metric functions have no peak in left panel, have peaks in the right panel.}
	\label{Fig30031}
\end{figure}

For $Q=1$, the metric functions with and without a peak are plotted in Figure~\ref{Fig30031}.

\begin{figure}[H]	
	\centering
	\includegraphics[width=0.75\textwidth]{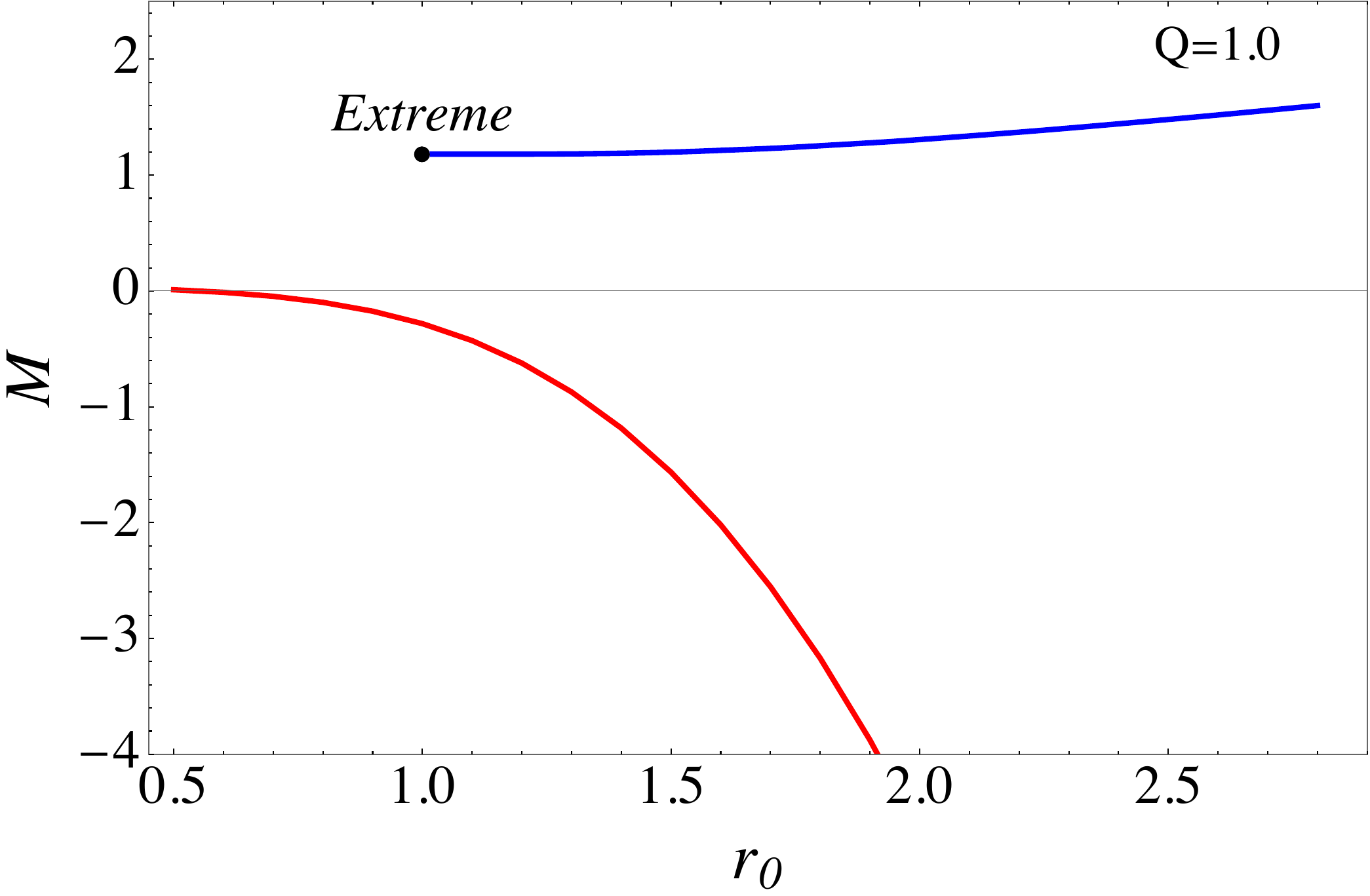}
	\caption{The mass is shown as a function of horizon radius for fixed charge $Q=1.0$. Due to the large charge $Q=1$, the charged solutions are far from the neutral solutions, thus the case for uncharged solutions is not included in this figure. }
	\label{Fig3004}
\end{figure}

The curves of mass versus horizon radius are shown as Figure~\ref{Fig3004}. From the mass plot, now it is clear that the "separation" between the two solutions becomes even more pronounced at this stage. 

\begin{figure}[H]	
	\centering
	\includegraphics[width=0.45\textwidth]{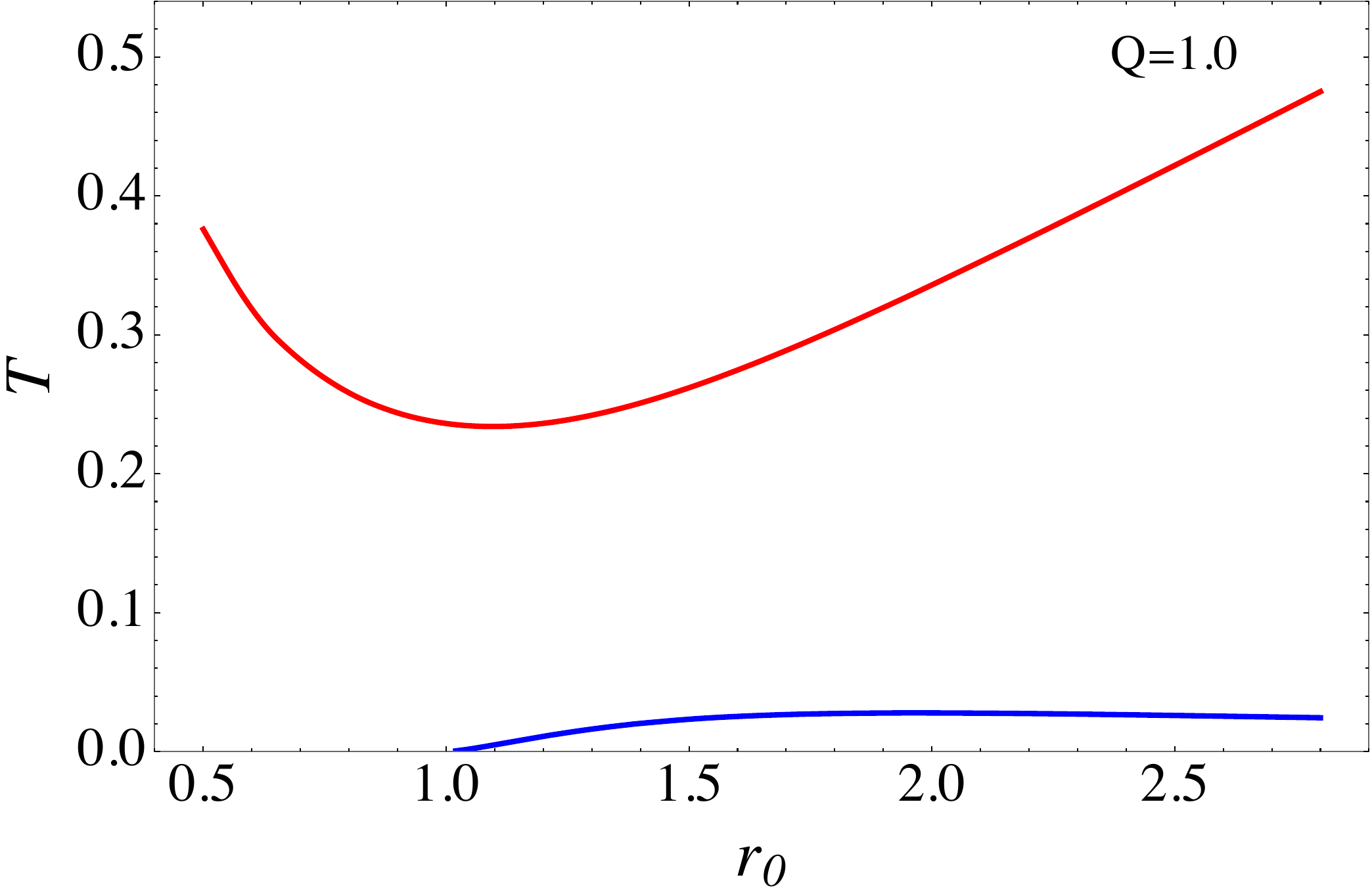}
	\qquad
	\includegraphics[width=0.45\textwidth]{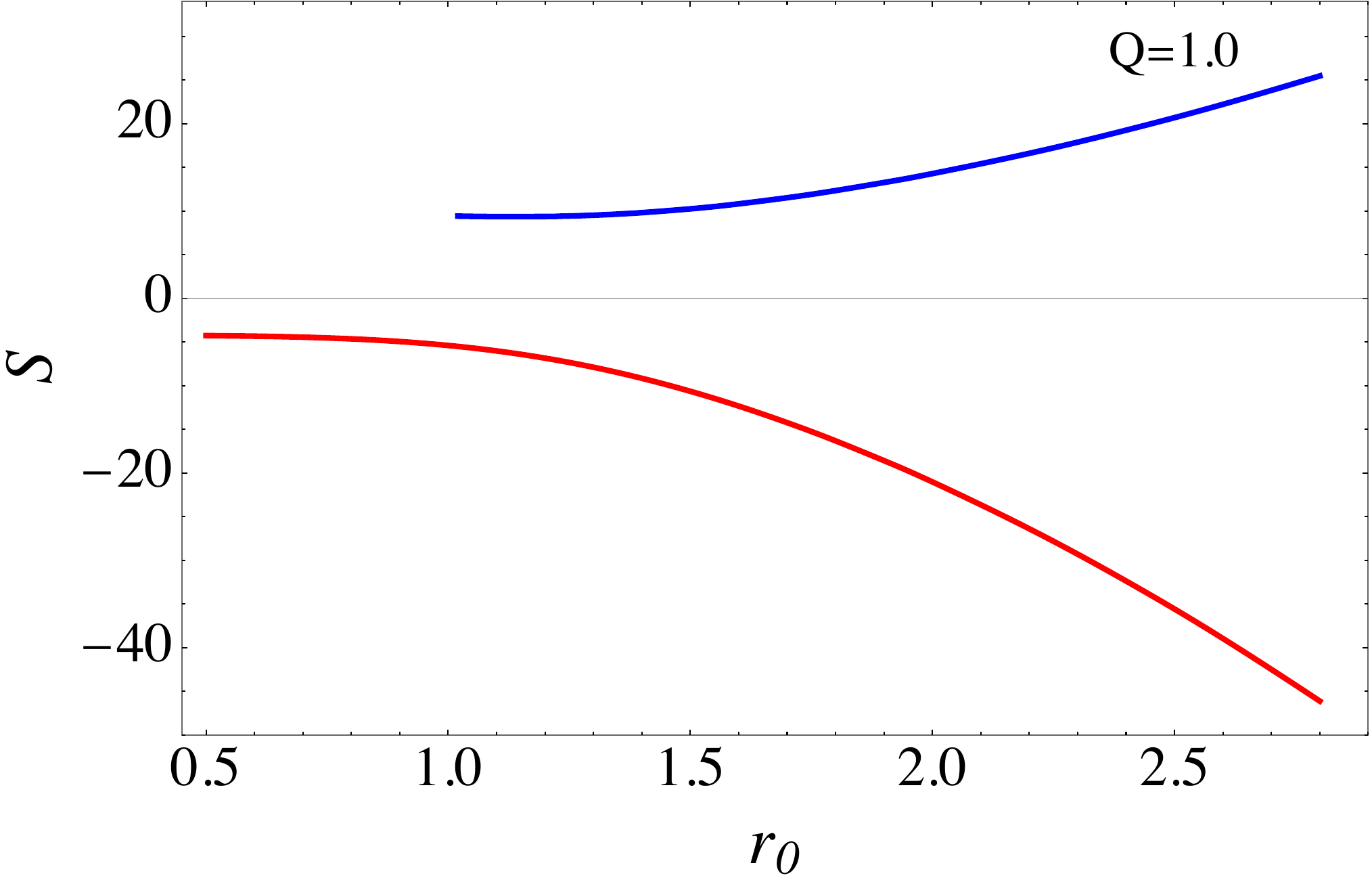}
	\caption{The temperature and entropy as functions of horizon radius is shown for fixed charge $Q=1.0$. }
	\label{Fig3003}
\end{figure}

The temperature and entropy as functions of  the event horizon radius are illustrated in the Figure~\ref{Fig3003}. As is seen, the charged black holes are far from each other. 

\begin{figure}[H]	
	\centering
	\includegraphics[width=0.75\textwidth]{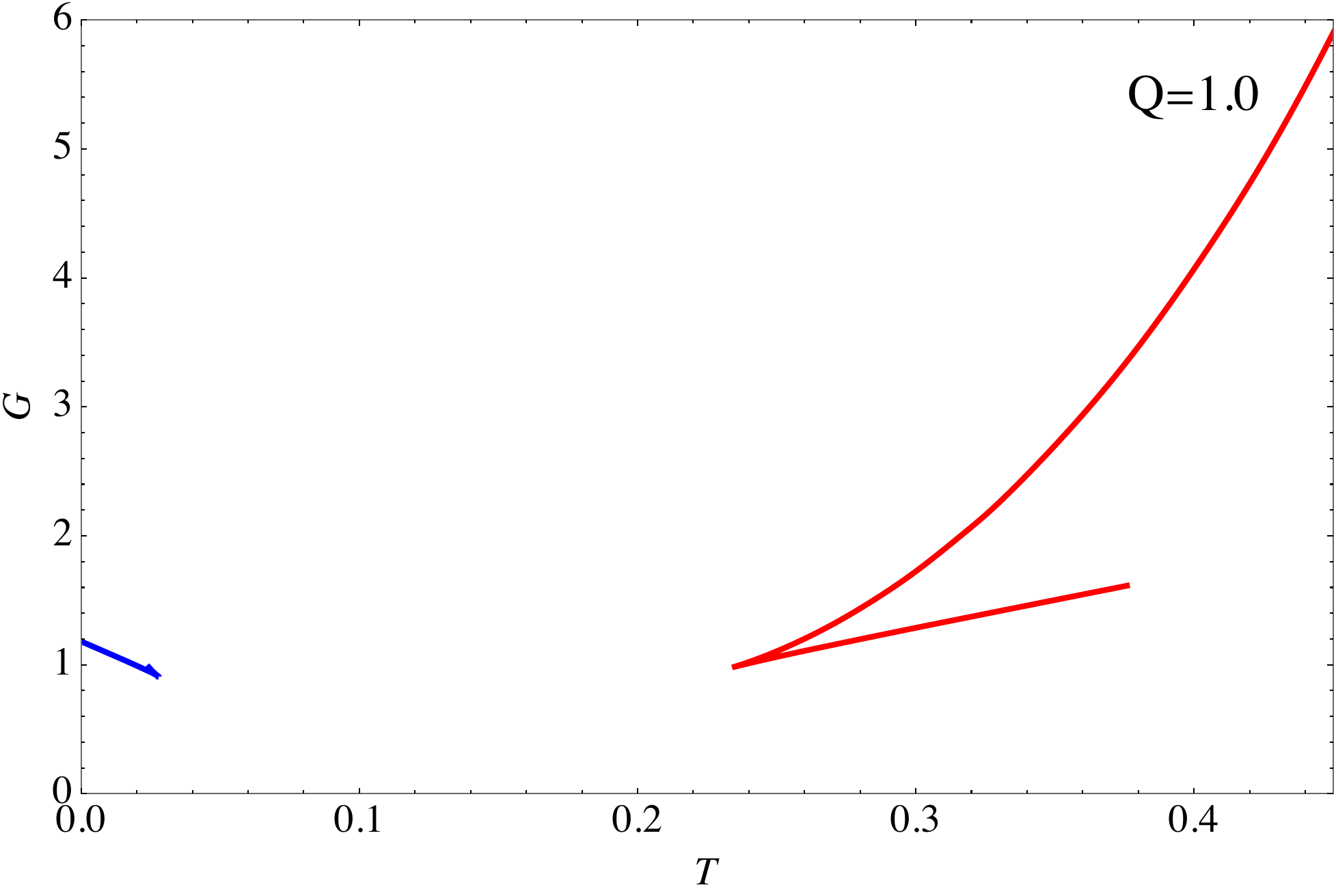}
	\caption{The  free energy $G$ as a function of temperature $T$ for $Q=1.0$, it is obvious that the two branches of charged solutions are far from each other.}
	\label{Fig30042}
\end{figure}

And the relationshp of temperature $T$ and free energy $G$ is depicted in the Figure~\ref{Fig30042}, the two branches of solutions are far from each other.

\begin{figure}[H]	
	\centering
	\includegraphics[width=0.75\textwidth]{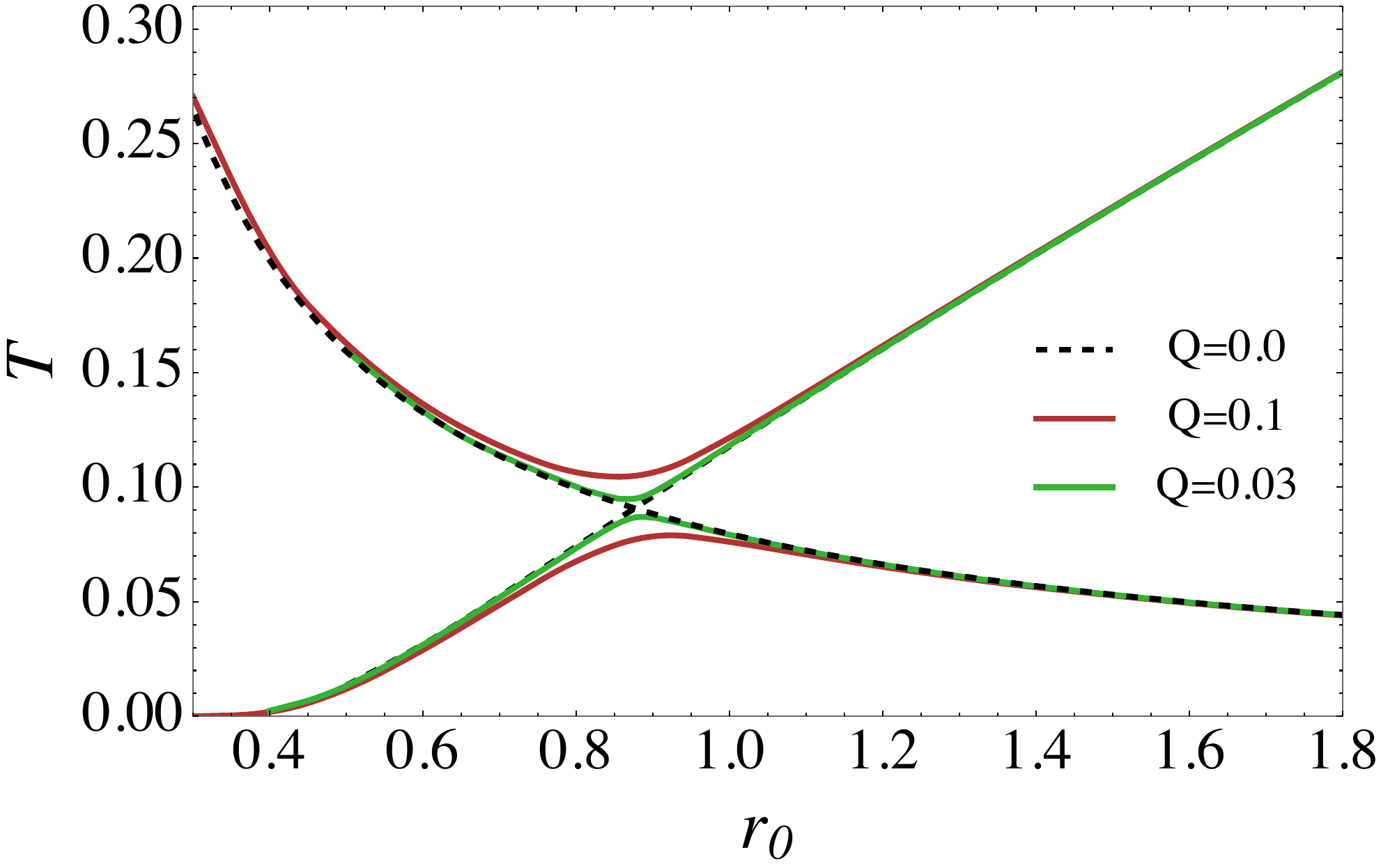}
	\caption{The temperature-radius relations for various values of charge $(Q = 0, 0.03, 0.1)$. The procedure that the charged solutions break off from the neutral ones is clearly shown. }
	\label{Fig30053}
\end{figure}

Finally, we want summarize this section with one picture including various charges. As the charge is increasing, the charged solutions goes far away from the neutral ones. The significant difference between the charged solutions and the neutral solutions is that the two neutral solutions intersect with each other whilst the two charged ones don't. Each charged solution seems to contain half of the Schwarzschild and half of the non-schwarzschild solution. Thus the charged solutions are not simple generalizations of the neutral ones, their properties have great difference. The black hole temperature profiles for different charge values $(Q = 0, 0.03, 0.1)$ are comparatively plotted in Figure~\ref{Fig30053}.

\section{The relations with RN black hole}\label{section 4}
It is known that the celebrated RN black hole is not a solution of Einstein-Weyl-Maxwell theory. However, as is shown, the Einstein-Weyl-Maxwell theory can admit two branches of charged black holes, then two questions emerge: (1) What is the relation between the numerical charged black holes and the RN black hole? (2) Why can't Einstein-Weyl-Maxwell theory support RN black hole? 

In order to compare the numerical charged black holes with RN black hole, we fix the charge and explore their temperature and horizon radius curves. We choose $Q=0.1$ to be the first example. As one can see, one of the charged black holes (solid blue line) is close to RN black hole (dashed black line), whilst the other charged black hole is totally different from RN black hole.

\begin{figure}[H]	
	\centering
	\includegraphics[width=0.75\textwidth]{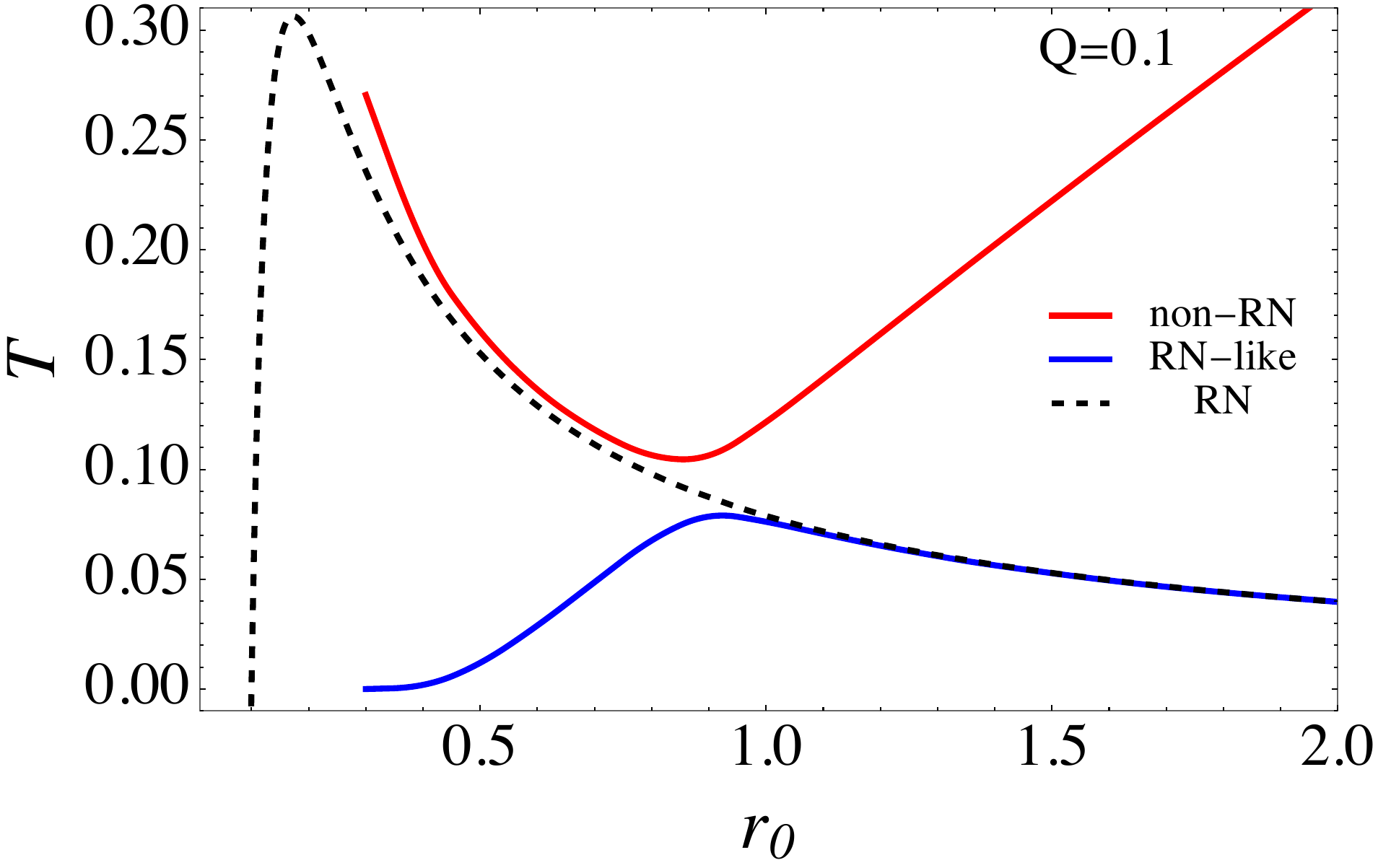}
	\caption{The red and blue lines show  the temperature of two branches of charged solutions for charge $Q=0.1$, the dashed line represents the temperature of a RN black hole for the same charge $Q=0.1$. One branch of the charged solution is similar to RN black hole, whilst the other branch is totally different from RN black hole. }
	\label{Fig3002}
\end{figure}

\begin{figure}[H]	
	\centering
	\includegraphics[width=0.44\textwidth]{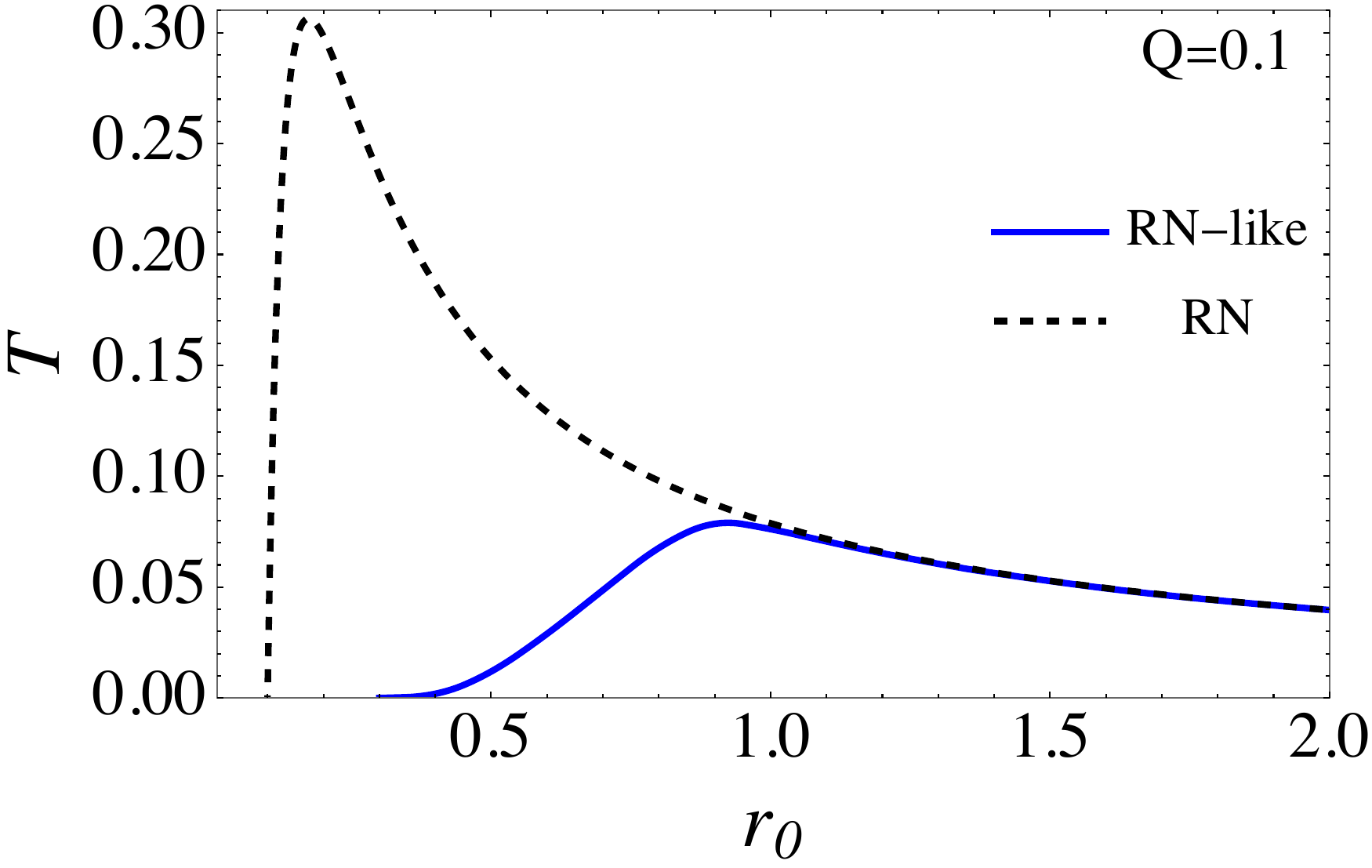}
	\qquad
	\includegraphics[width=0.44\textwidth]{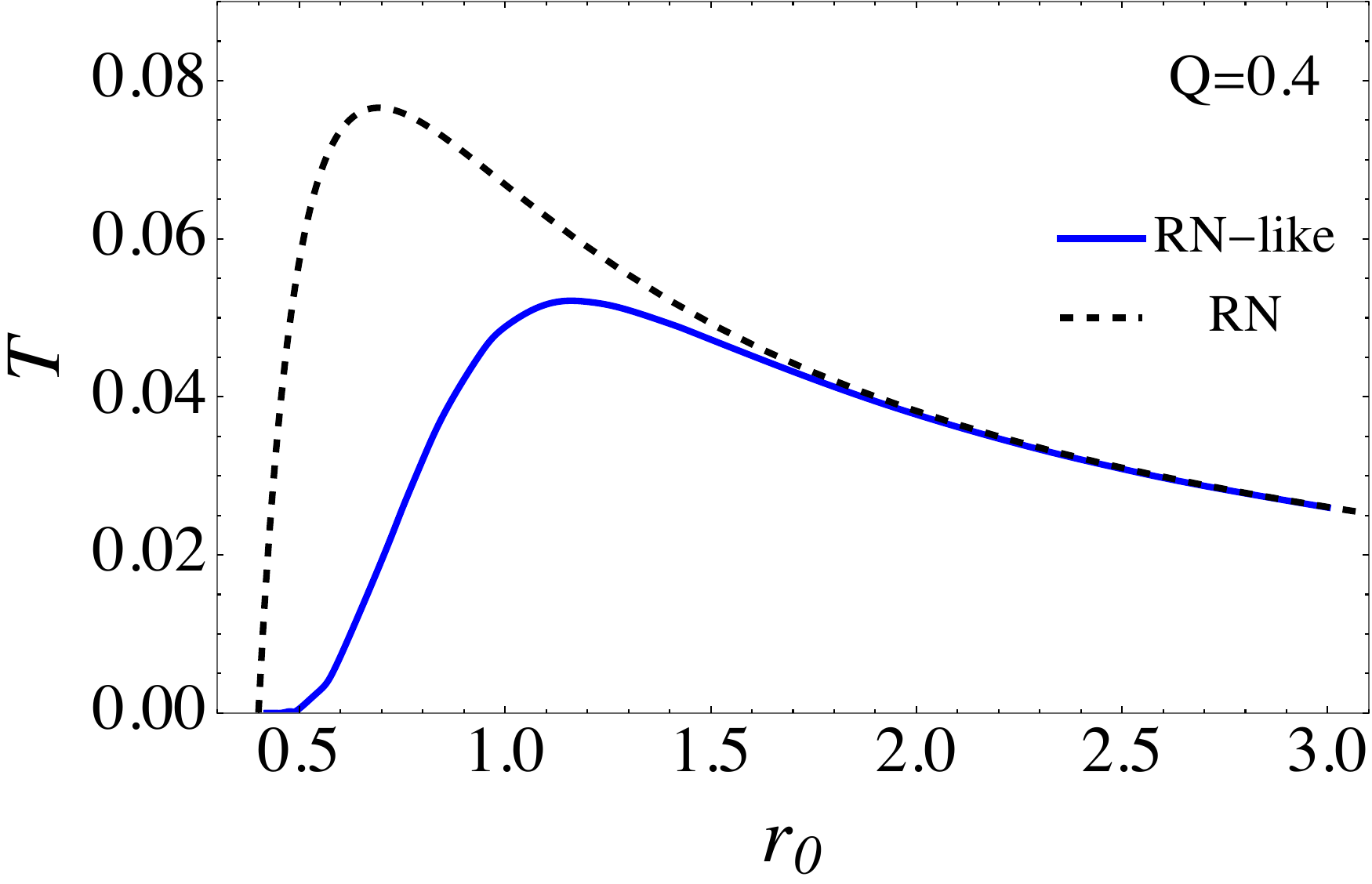}
	\qquad
	\includegraphics[width=0.44\textwidth]{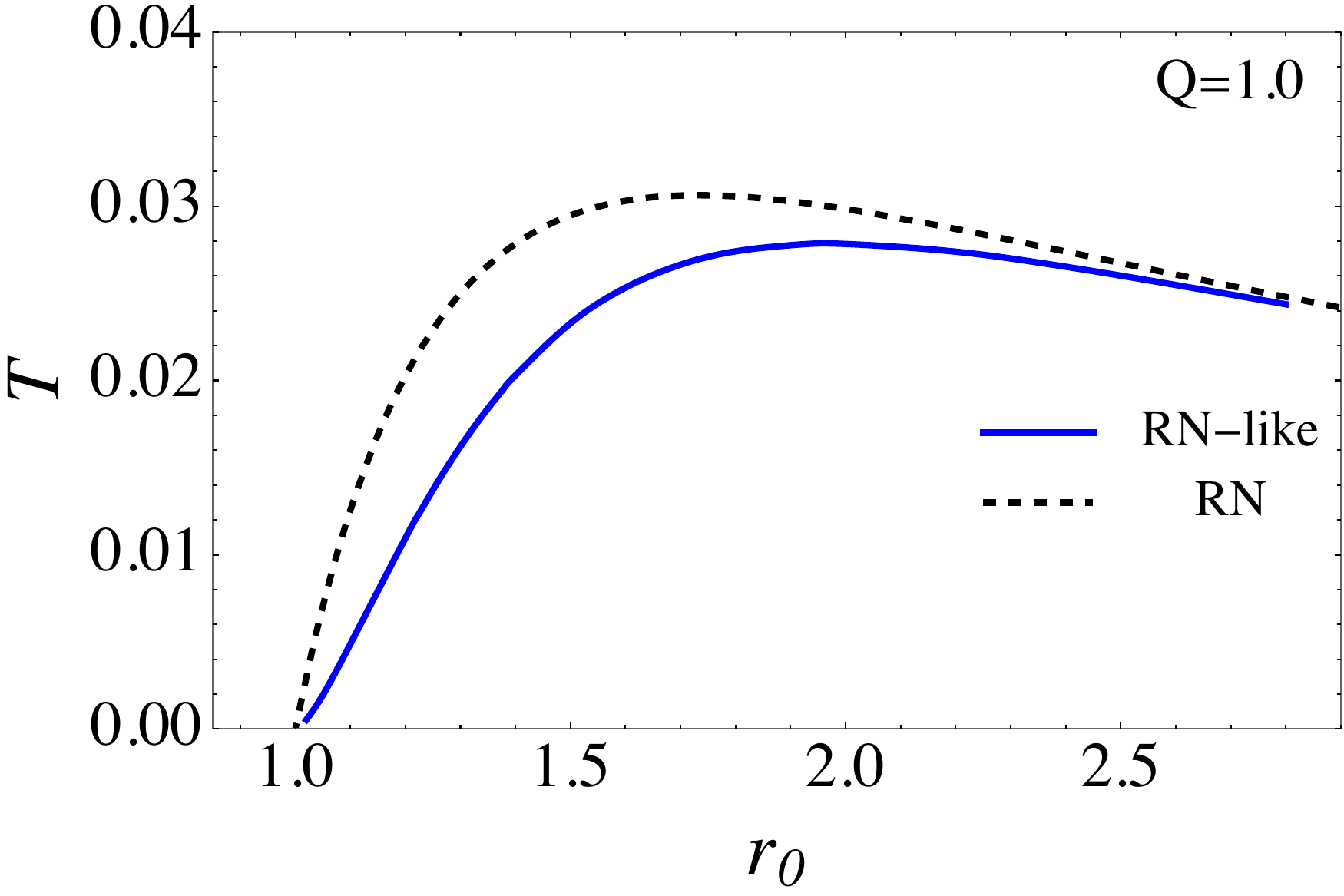}
	\qquad
	\includegraphics[width=0.45\textwidth]{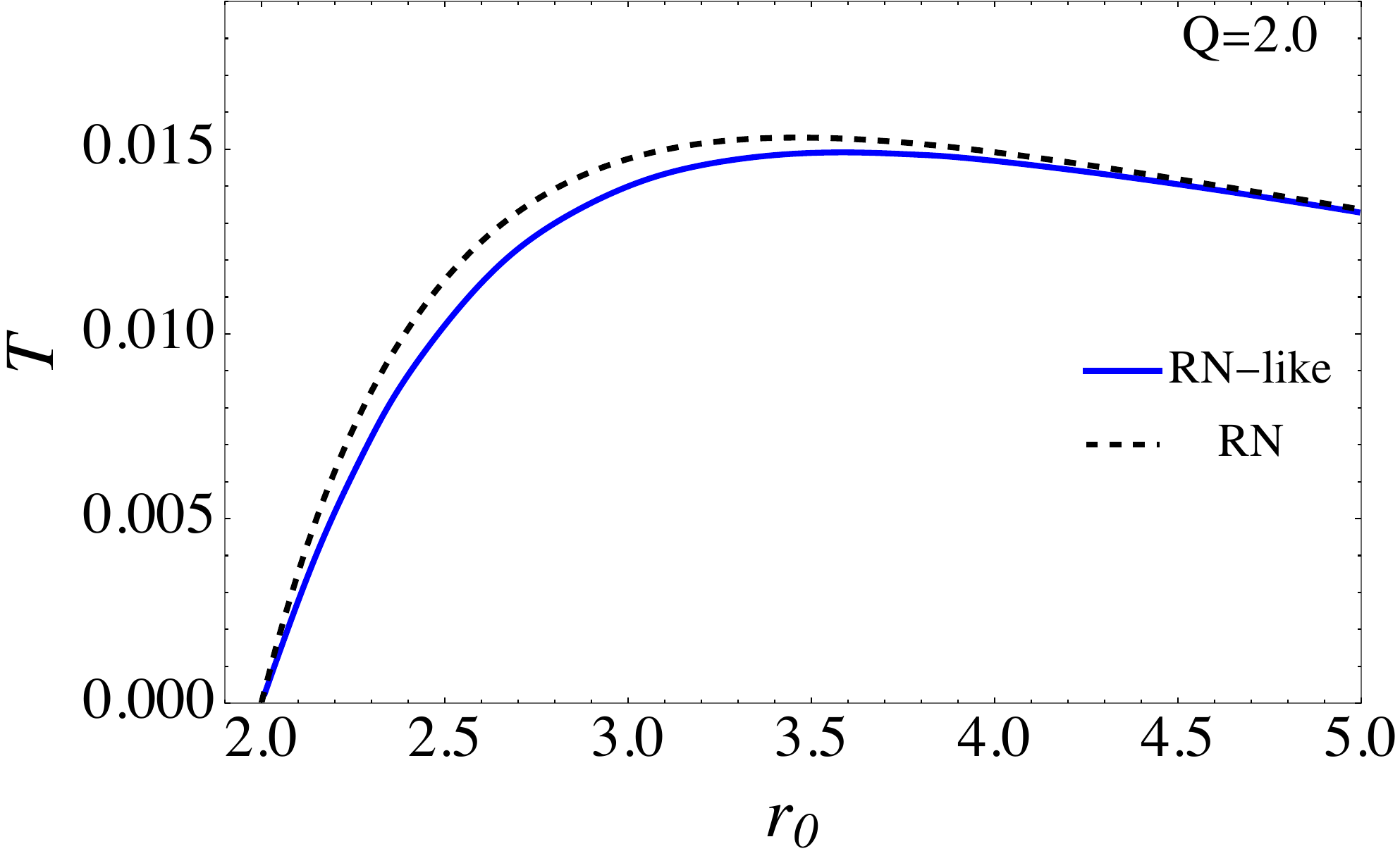}
	\caption{The temperature of RN-like black hole as a function of horizon radius is  shown for various charges $Q=0.1, 0.4, 1, 2$. The RN-like black hole approaches RN black hole as charge increases. }
	\label{Fig30051}
\end{figure}

Then we explore more values of charge, the $T-r_0$ curves are shown in Figure~\ref{Fig30051}. We observe that one of charged black hole is totally different from the RN black hole, and thus we ignore this branch of charged black hole in Figure~\ref{Fig30051}. The other branch of solution is similar to RN black hole. For this reason, we shall call the branch which is totally different from RN black hole "non-RN" black hole, and the one which is similar to RN black hole "RN-like" black hole.  As the charge is increasing, the charged black hole is approaching the RN black hole.

\begin{figure}[H]	
	\centering
	\includegraphics[width=0.94\textwidth]{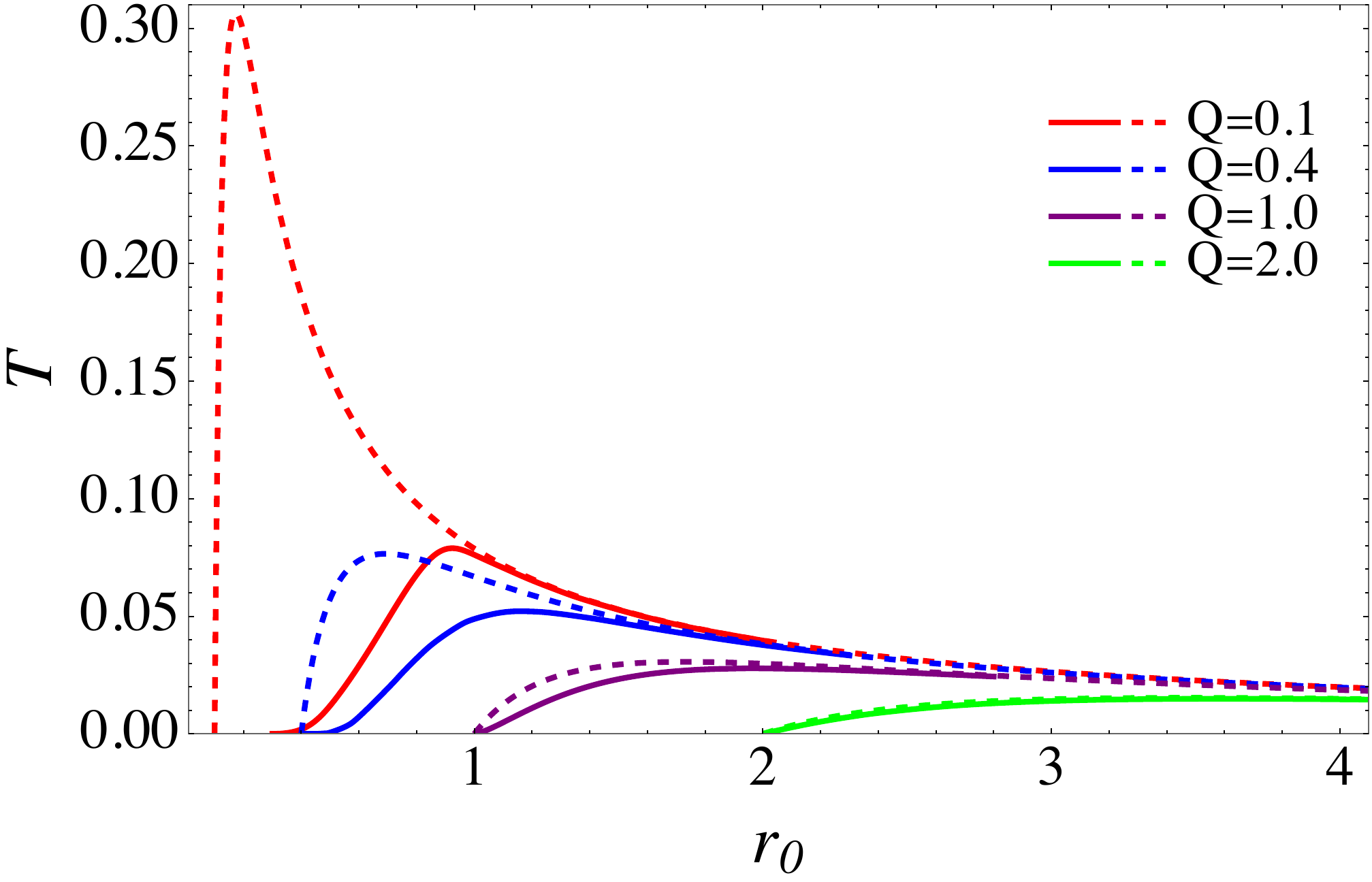}
	\caption{The temperature plots for different charges are combined into a single figure, clearly illustrating the tendency that the RN-like black hole approaches the RN black hole as the charge increases.}
	\label{Fig30052}
\end{figure}

We put $T-r_0$ curves for various charges in one figure in Figure~\ref{Fig30052}.

\section{The first law and stability of the charged black holes}\label{section 5}
\subsection{The first law}\label{section 5.1}
\begin{figure}[H]	
	\centering
	\includegraphics[width=0.45\textwidth]{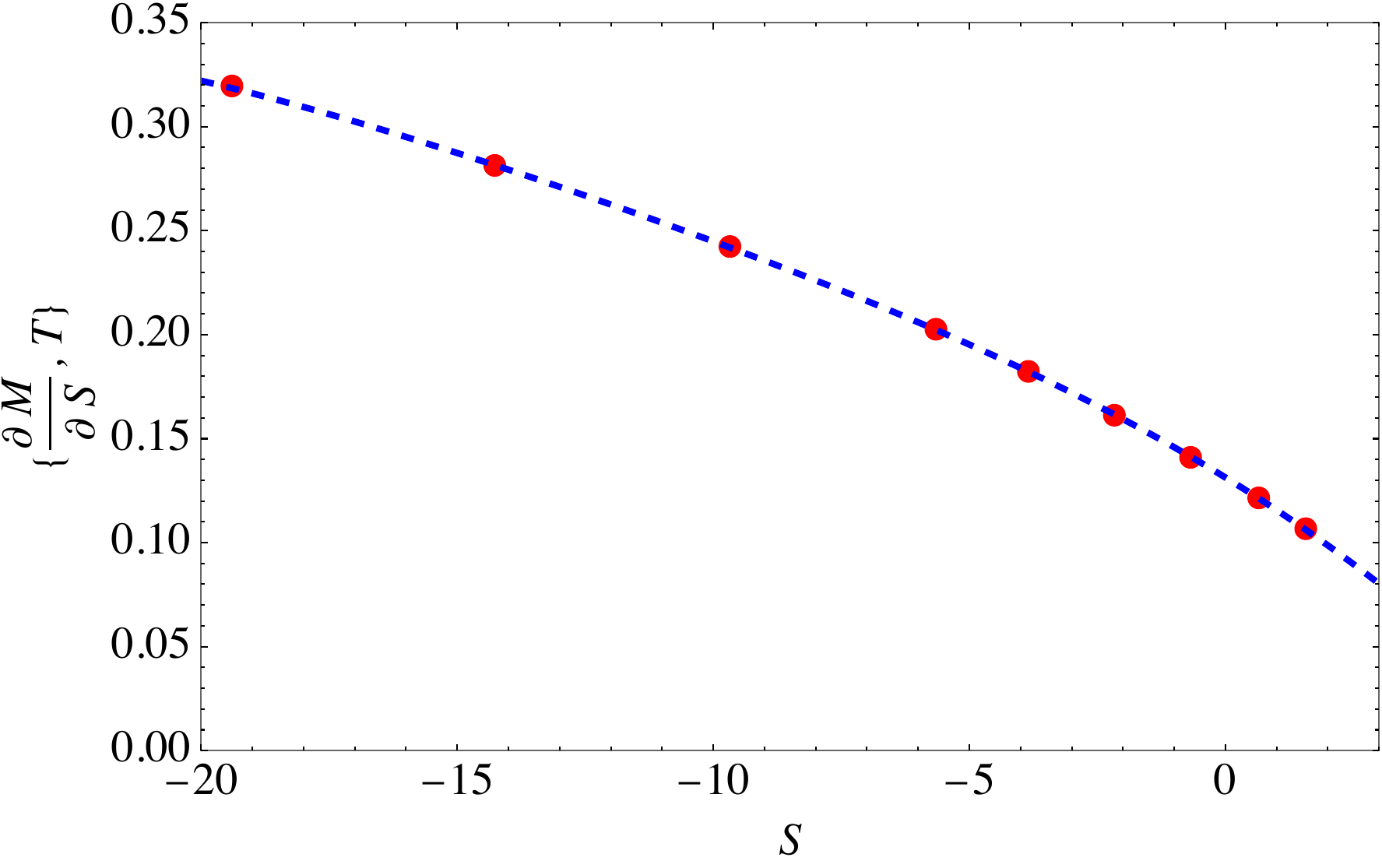}
	\qquad
	\includegraphics[width=0.45\textwidth]{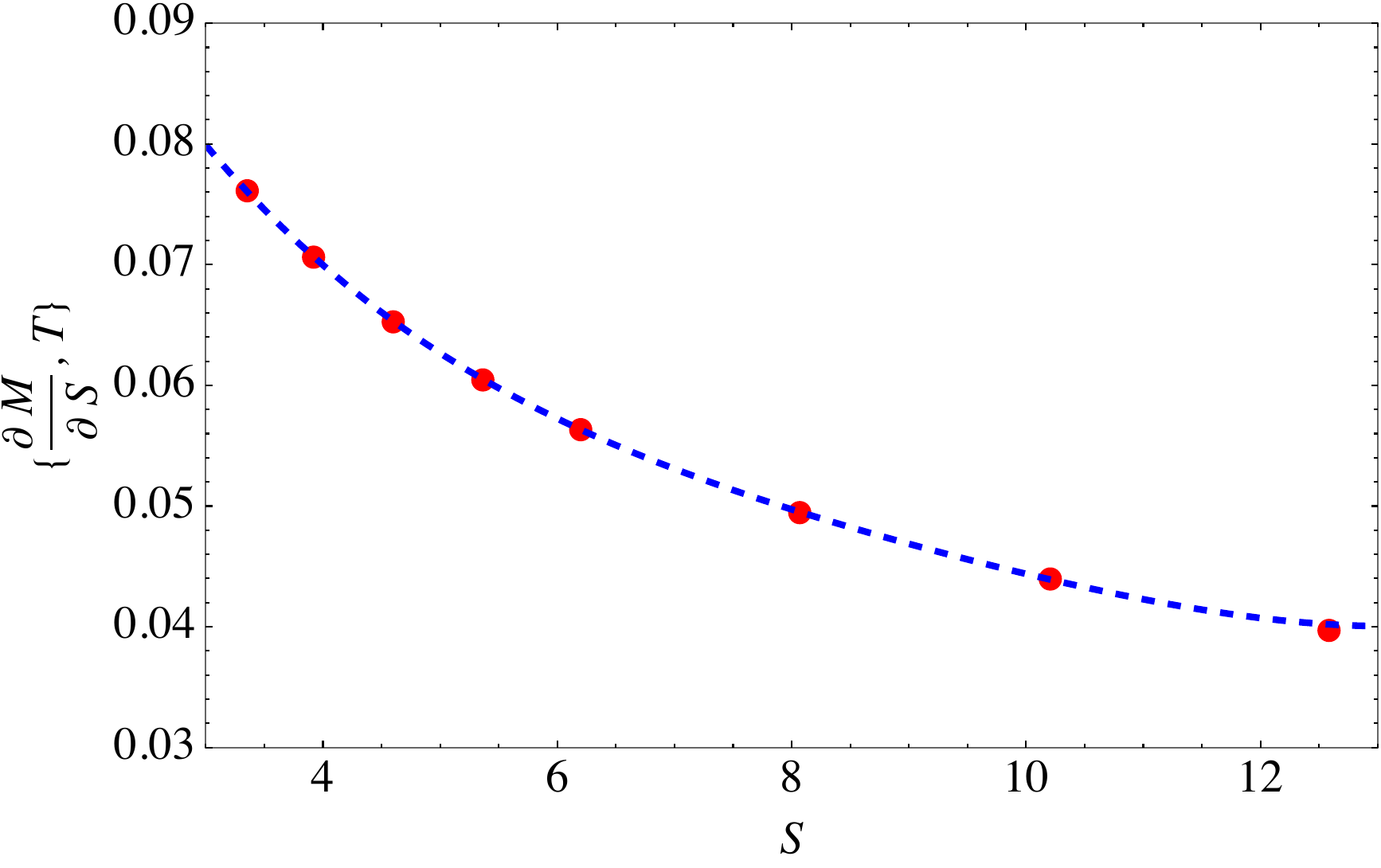}
	\caption{The blue dashed line represents $ \left(\partial M / \partial S\right)_{Q} $, the red dots denote the temperature $T$. Left panel correspond the "non-RN" black holes, the right panel correspond the "RN-like" black holes.}
	\label{Fig40051}
\end{figure}

Verifying the thermodynamic first law of the numerical black hole solutions is tricky, especially for charged black holes which have two or more independent parameters. In this section, we address in confirming the first law by generalizing the method in~\cite{Lu:2015cqa}. The first law of charged black hole has the form 
\begin{equation}
dM = T dS + \Phi dQ \,.
\end{equation}
Thus, the following Maxwell relations must hold
\begin{equation}
	\label{eq40051}
	\left(\frac{\partial M}{\partial S}\right)_{Q}=T, \qquad \left(\frac{\partial M}{\partial Q}\right)_{S}=\Phi.
\end{equation}
We first set the charge to be $Q=0.1$ and fit the mass  as a function of entropy $M(S)$. Then we can obtain the  $ \left(\partial M / \partial S\right)_{Q} $ through the fitting function $M(S)$. The first Maxwell relation can be confirmed by comparing $ \left(\partial M / \partial S\right)_{Q} $ with the temperature $T$. As is shown in Figure~\ref{Fig40051}, the blue dashed line represents $ \left(\partial M / \partial S\right)_{Q} $, whilst the red dots denote the temperature \( T \).  The left and right panels correspond to two different branches of the charged black hole solutions. It can be seen from the figure that the $ \left(\partial M / \partial S\right)_{Q} $ and $T$ match in a great accuracy. 

\begin{figure}[H]	
	\centering
	\includegraphics[width=0.45\textwidth]{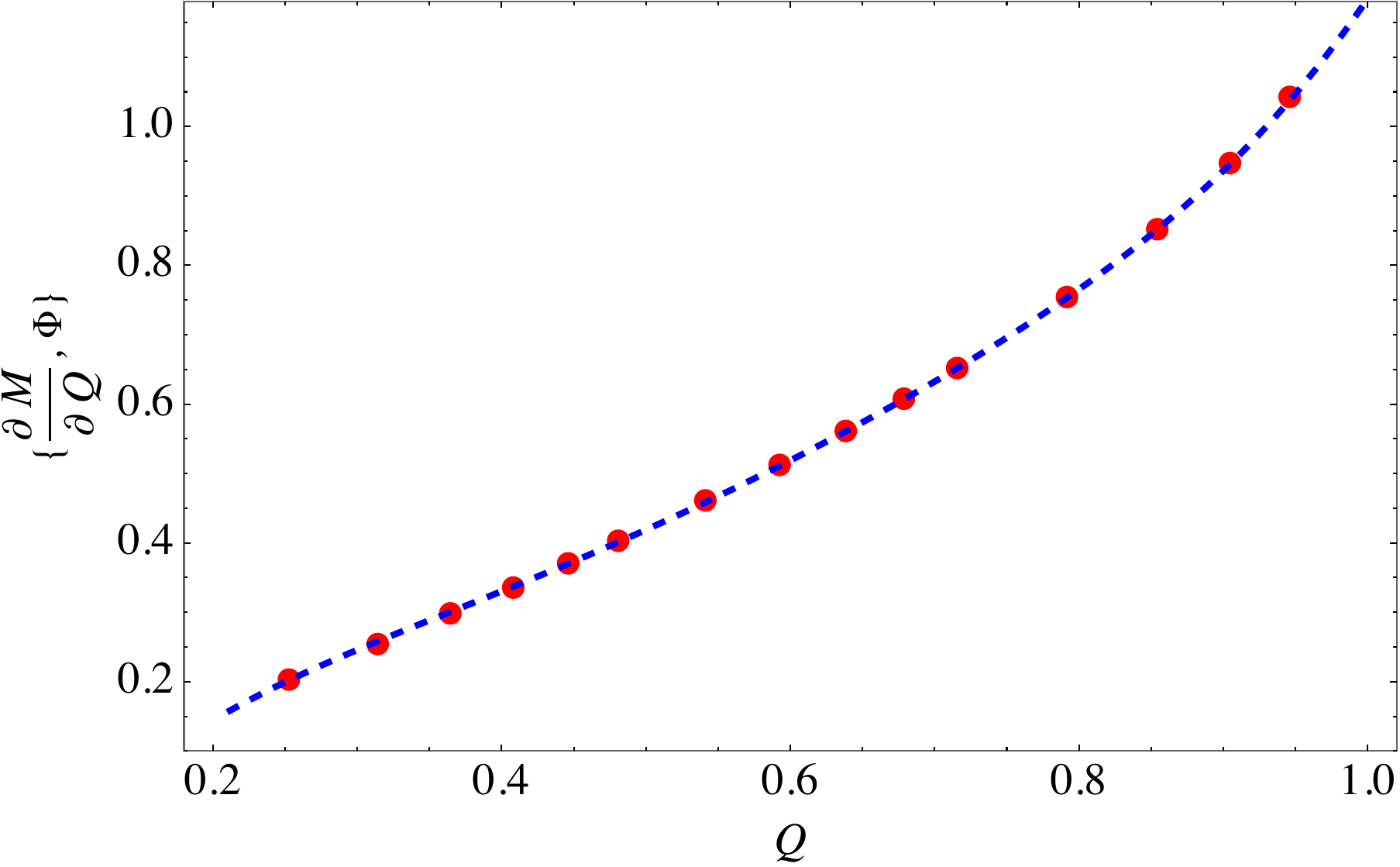}
	\qquad
	\includegraphics[width=0.45\textwidth]{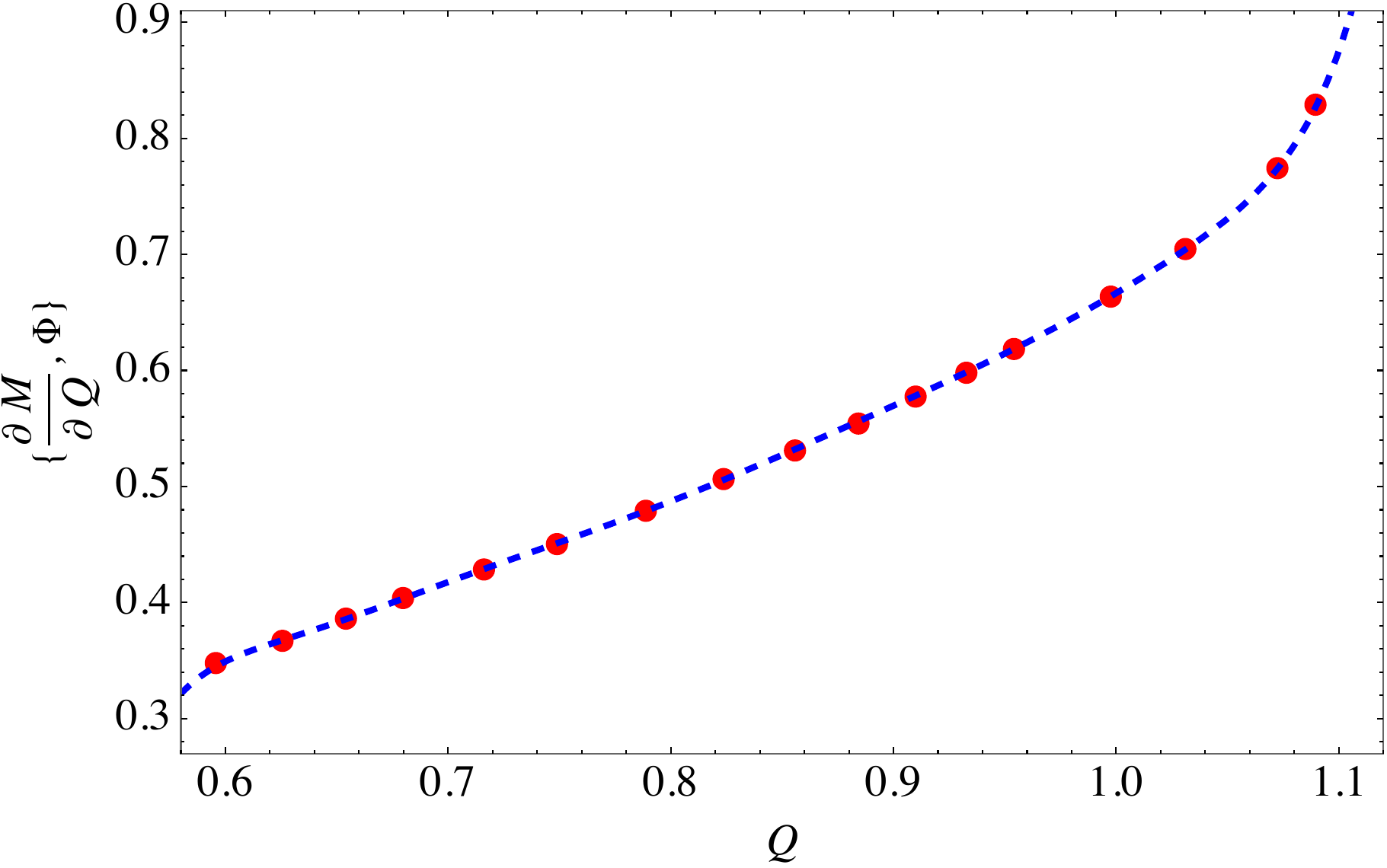}
	\caption{The blue dashed line represents $ \left(\partial M / \partial Q\right)_{S} $, the red dots denote the electric potential $\Phi$. Left panel correspond the "non-RN" black holes, the right panel correspond the "RN-like" black holes.}
	\label{Fig40052}
\end{figure}

The second Maxwell relation can be confirmed through the same procedure. This time, we fix the entropy \( S \) and fit the mass as a  function of charge \( M(Q) \). $ \left(\partial M / \partial Q\right)_{S} $ can be directly calculated through the fitting function  \( M(Q) \). We compare $ \left(\partial M / \partial Q\right)_{S} $ with electric potential $\Phi$ in Figure~\ref{Fig40052}. The left and right panels  correspond to two different branches of the charged solutions. Again, the  $ \left(\partial M / \partial Q\right)_{S} $ totally matches electric potential $\Phi$ for both branches of hairy charged black holes. 
To this end, the two Maxwell relations hold exactly and the first law is confirmed.

\subsection{Stability Analysis of Black Hole Solutions via Quasinormal Modes}\label{section 5.2}
In this chapter, we investigate the stability of our black hole solution by analyzing its quasinormal modes (QNMs), which characterize the characteristic vibrational response of a black hole to external perturbations (e.g., gravitational waves, matter perturbations). For a stable black hole, these perturbations decay over time through damped oscillations, termed quasinormal modes (QNMs). The complex frequencies $\omega=\omega_R+ i \omega_I$ of these modes encode crucial stability information: a negative imaginary part $(\omega_I < 0)$ signifies exponential decay of the perturbations, indicating stability, whereas a positive imaginary part $(\omega_I > 0)$ would imply exponential growth and instability.

To compute the quasinormal frequencies $\omega$ for our solution, we employ two methods. First, the finite difference method is applied to solve the time-evolution equation numerically, offering high precision for general black hole spacetimes. Second, the higher-order WKB approximation is utilized to validate results and explore analytical insights. By comparing results from these methods, we rigorously assess the stability properties of the black hole solution under consideration.

We consider the evolution of a massless scalar field $\Phi(t,r,\theta,\phi)$ as a perturbation on the background of our black hole solution, governed by the Klein-Gordon equation in curved spacetime. Given that the spacetime is stationary and spherically symmetric, we decompose the scalar field as follows:
\begin{equation}
	\Phi(t,r,\theta,\phi) =  \sum_{l=0}^{\infty} \sum_{m=-l}^{l} \frac{\phi(t,r)}{r}  Y_{lm}(\theta,\phi),
	\label{eq:scalar_decomposition}
\end{equation}
where \( Y_{lm}(\theta,\phi) \) are spherical harmonics. Substituting this decomposition into the Klein-Gordon equation yields
\begin{equation}
	- \frac{1}{h(r)} \frac{\partial^2 \phi}{\partial t^2} + \frac{1}{r^2}\sqrt{\frac{f(r)}{h(r)}} \frac{\partial}{\partial r}\left(r^2 \sqrt{h(r) f(r)} \frac{\partial \phi}{\partial r}\right) - \frac{l(l+1)}{r^2} \phi = 0,
	\label{eq:kg_equation}
\end{equation}
where \( l \) is the angular momentum quantum number.

To simplify the radial coordinate dependence, we introduce the tortoise coordinate $ r_* $, defined such that
\begin{equation}
	\frac{dr_*}{dr} = \frac{1}{\sqrt{h(r) f(r)}}.
	\label{eq:tortoise_coord}
\end{equation}
Under this transformation, Equation~\eqref{eq:kg_equation} reduces to the standard Schr\"{o}dinger-like form:
\begin{equation}
	\left[\frac{\partial^2 }{\partial t^2} - \frac{\partial^2 }{\partial r_*^{2}} + V(r)\right] \phi(r,t) = 0,
	\label{eq:kg_tortoise}
\end{equation}
with
\begin{equation}
	V(r)=\frac{l(l+1) h(r)}{r^2}+ \frac{\partial_{r} \left(h(r) f(r)\right)}{2r},
	\label{eq:Vr_effpot}
\end{equation}
where \( V(r) \) is an effective potential which explicitly incorporates the geometric effects of the black hole spacetime.

We employ the numerical method proposed in Reference~\cite{Gundlach:1993tp} to solve Equation~\eqref{eq:kg_tortoise}. To facilitate the calculations, we introduce the light-cone coordinates $u=t-r_*$ and $v=t+r_*$. Under this transformation, Equation~\eqref{eq:kg_tortoise} takes the form of 
\begin{equation}
	\frac{\partial^2 }{\partial u \partial v} \phi(u,v) + \frac{1}{4}V \phi(u,v) = 0,
	\label{eq:kg_uv}
\end{equation}
By discretizing the $uv-$plane into a grid, we can express Equation~\eqref{eq:kg_uv} using finite difference methods
\begin{equation}
	\begin{aligned}
		\phi(u + \Delta u,v + \Delta v) = &\phi(u,v + \Delta v) + \phi(u + \Delta u,v) - \phi(u,v)\\
		&-\frac{\Delta u \Delta v}{8} V(u,v) \left[\phi(u,v + \Delta v) + \phi(u + \Delta u,v)\right],
		\label{eq:kg_dudv}
	\end{aligned}
\end{equation}
where $\Delta u$ and $\Delta v$ represent the step sizes. For the initial conditions, we impose a Gaussian-type perturbation 
\begin{equation}
	\phi(u_0,v) = A \text{e}^{\frac{(v-v_c)^2}{\sigma_{v}^2}},\qquad \partial_v \phi(u_0,v) =0.
	\label{eq:initial_Gaussian}
\end{equation}
at the $u=u_0$ boundary. Here $A$, $v_{c}$ and $\sigma_{v}$ are tunable parameters that do not influence the outcome of $\omega$. In the subsequent calculations, we set $l=6$, and the potential $V(r)$ is given by Equation~\eqref{eq:Vr_effpot}. With these parameters, we could numerically solve for $\phi(u,v)$. 

\begin{figure}[H]	
	\centering
	\includegraphics[width=0.75\textwidth]{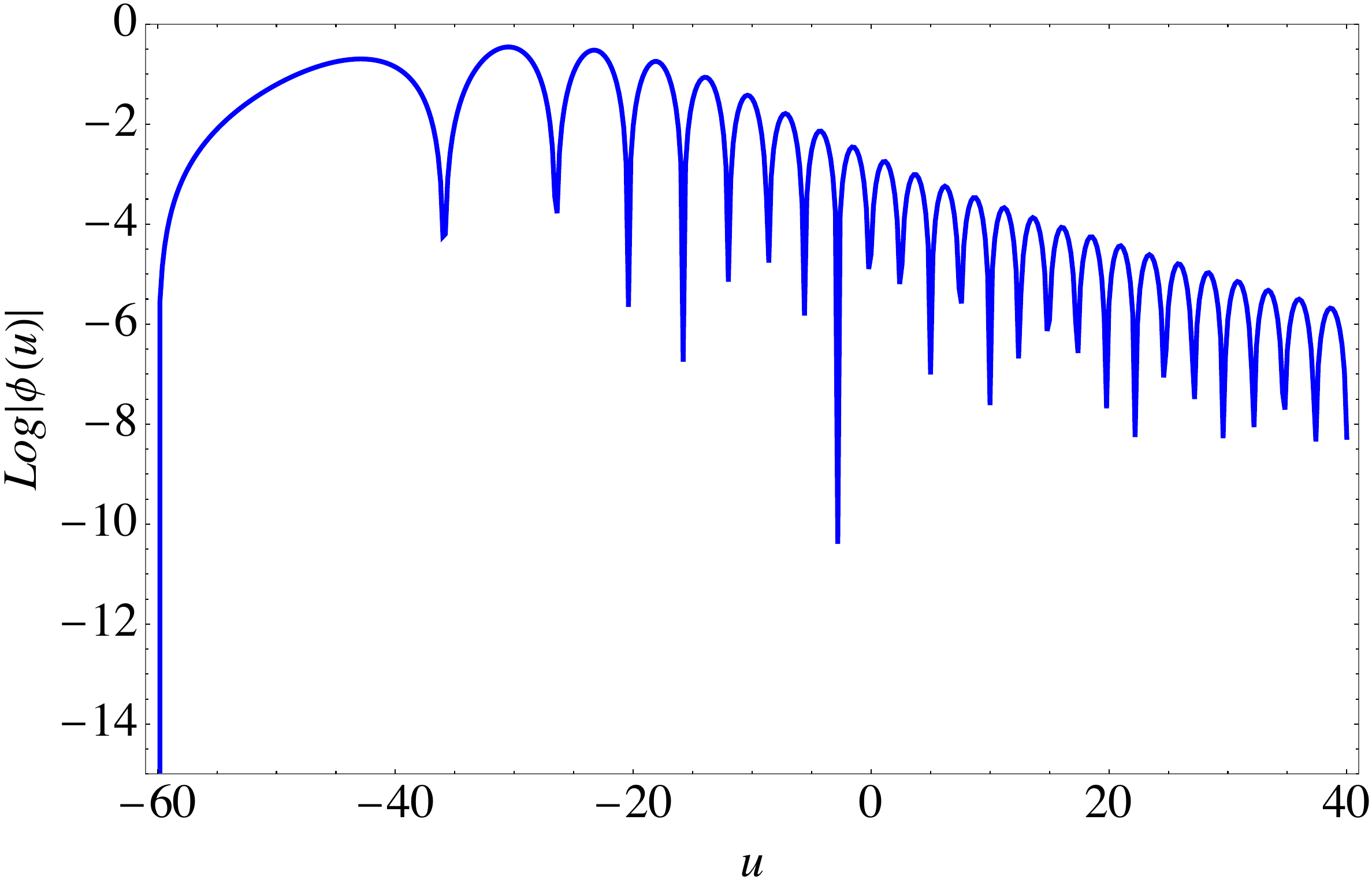}
	\caption{The plot of $u-Log{|\phi(u)|}$, with $Q=1.0, r_{0}=1.4$, and the angular momentum quantum number $l=6$.}
	\label{Fig40001}
\end{figure}

After obtaining the numerical solution $\phi(u,v)$, the subsequent task is to extract the quasinormal frequencies $\omega_i$. By selecting $\phi(u,v)$ at sufficiently large values of $v=v_{large}$, we expect the behavior to follow 
\begin{equation}
	\phi(u)\big|_{v=v_{large}} \sim \sum_{i=0} C_{i} \text{e}^{- i \omega_i u},
	\label{eq:phi_v_large}
\end{equation}
where $\omega_i$ corresponds to the quasinormal frequencies of different excitation states. Figure~\ref{Fig40001} shows the logarithmic plot of the function 
$\phi(u)\big|_{v=v_{large}}$, the waveform exhibits distinct damped oscillations. We employ the Prony method to extract $\omega_i$, with a particular focus on the fundamental mode $\omega_0$. The results are summarized in the Table~\ref{tab:omega_0_I} and Table~\ref{tab:omega_0_II}.

Next, we employ the WKB approximation method to obtain the quasinormal frequencies $\omega$. This method is widely used in quantum mechanics to analyze the stationary Schrödinger equation. By expressing the scalar field $\phi(t,r)$ as $\phi(t,r)=\text{e}^{-i \omega t} R(r)$, where is $R(r)$ is a function of spatial coordinates. We substitute this ansatz into the Equation~\eqref{eq:kg_tortoise}. This results in the separated equation:
\begin{equation}
	\frac{\partial^2 R(r)}{\partial r_*^{2}} + \left[\omega^2 - V(r)\right] R(r) = 0,
	\label{eq:kg_tortoise-r}
\end{equation}
By applying this approach, we obtain an equation analogous to the stationary Schrödinger-like equation. Following the third-order WKB approximation method outlined in reference~\cite{Iyer:1986np}, the quasinormal frequencies are derived
\begin{equation}
	\omega^2 = V_0 - i \sqrt{-2 V_0''} \left[\left( n + \frac{1}{2}\right) + \Lambda_{2} +\Lambda_{3}\right],
	\label{eq:3th_WKB_omega}
\end{equation}
where $\Lambda_{2}$ and $\Lambda_{3}$ are
\begin{equation}
	\begin{aligned}
		\Lambda_{2} & =\frac{1}{\sqrt{-2 V_0''}} \left[ \frac{1}{8} \frac{V_0^{(4)}}{V_0''} \left( \frac{1}{4} + \left(n + \frac{1}{2}\right)^2 \right) - \frac{1}{288} \left( \frac{V_0'''}{V_0''} \right)^2 \left( 7 + 60 \left(n + \frac{1}{2}\right)^2 \right) \right],
		\\
		\Lambda_{3} & =\frac{1}{-2 V_0''} \left( n + \frac{1}{2} \right)  \left[ \frac{5}{6912} \left(\frac{V_0'''}{V_0''}\right)^4\left(77+188\left(n + \frac{1}{2}\right)^2\right) \right.
		\\
		&\left. -\frac{1}{384} \left(\frac{V_0'''^2 V_0^{(4)}}{V_0''^3}\right)\left(51+100\left(n + \frac{1}{2}\right)^2\right) +\frac{1}{2304} \left(\frac{V_0^{(4)}}{V_0''}\right)^2\left(67+68\left(n + \frac{1}{2}\right)^2\right) \right.
		\\
		&\left.
		+\frac{1}{288} \left(\frac{V_0''' V_0^{(5)}}{V_0''^2}\right)\left(19+28\left(n + \frac{1}{2}\right)^2\right)
		-\frac{1}{288} \left(\frac{ V_0^{(6)}}{V_0''}\right)\left(5+4\left(n + \frac{1}{2}\right)^2\right) \right].
	\end{aligned}
	\label{eq:3th_WKB_Lambda_2_3}
\end{equation}
In this context, $V_{0}$ represents the maximum value of the potential $V(r_{*})$, $V_{0}^{(n)}$ denotes the $n-$th derivative of the potential function $V(r_{*})$ evaluated at its extremum point $r_{* ext}$, which satisify $V'(r_{* ext})=0$. Using Equation~\eqref{eq:3th_WKB_omega}, and setting $l=6$ and $n=0$, we could compute the values of $\omega_0$, and the results are summarized in Table~\ref{tab:omega_0_I} and Table~\ref{tab:omega_0_II}.

\begin{table}[H]
	\centering
	\begin{tabular}{cc|cc}	
		\hhline{====}
		$Q$ & $r_{h}$ & $\omega_0$-Prony-RN-like & $\omega_0$-WKB-RN-like  \\
		\hline
		0.03& 0.7 & $2.59842 - 0.150047 i$ & $2.61996 - 0.135494 i$ \\
		0.03& 0.9 & $2.76449 - 0.207555 i$ & $2.77597 - 0.189035 i$ \\
		0.1 & 0.6 & $2.39248 - 0.157417 i$ & $2.45686 - 0.112412 i$ \\
		0.1 & 0.8 & $2.66344 - 0.173350 i$ & $2.67199 - 0.157401 i$ \\
		0.1 & 0.9 & $2.64811  - 0.190205 i$ & $  2.65860 - 0.173847 i$ \\
		0.1 & 1.0 & $2.45944 - 0.183821 i$ & $2.47248 - 0.166828 i$ \\
		0.1 & 1.2 & $2.07678 - 0.164197 i$ & $  2.08613 - 0.142617 i$ \\
		0.1 & 1.4 & $1.77521 - 0.137505 i$ & $1.79360 - 0.123104 i$ \\
		1.0 & 1.02& $1.21834 - 0.057885 i$ & $1.33379 - 0.064091 i$ \\
		1.0 & 1.4 & $1.28804 - 0.075773 i$ & $1.31014 - 0.067138 i$ \\
		\hhline{====}
	\end{tabular}
	\caption{The quasinormal frequencies $\omega_0$ of the "RN-like" black holes, with $l=6$ and $n=0$. Here, $\omega_0$-Prony-RN-like denotes the quasinormal frequencies obtained using the finite difference method, while $\omega_0$-WKB-RN-like represents those calculated via the third-order WKB approximation.}
	\label{tab:omega_0_I}
\end{table}
\begin{table}[H]
	\centering	
	\begin{tabular}{cc|cc}
		\hhline{====}
		$Q$ & $r_{h}$ & $\omega_0$-Prony-non-RN & $\omega_0$-WKB-non-RN \\
		\hline
		0.03& 0.7 & $3.60055 - 0.271044 i$ & $3.60644 - 0.247112 i$ \\
		0.03& 0.9 & $2.91855 - 0.231725 i$ & $2.93137 - 0.209805 i$ \\
		0.1 & 0.6 & $4.26080 - 0.320788 i$ & $4.25762 - 0.293485 i$ \\
		0.1 & 0.8 & $3.24096 - 0.252440 i$ & $3.25135 - 0.229207 i$ \\
		0.1 & 0.9 & $3.03471 - 0.251257 i$ & $  3.04884 - 0.225957 i$ \\
		0.1 & 1.0 & $3.05558 - 0.292616 i$ & $3.07177 - 0.251944 i$ \\
		0.1 & 1.2 & $3.2374 - 0.361393 i$ & $3.21793 - 0.287676 i$ \\
		0.1 & 1.4 & $3.29834 - 0.472781 i$ & $3.34752 - 0.386779 i$ \\
		1.0 & 1.0 & $4.33777 - 0.542382 i$ & $4.35771 - 0.453357 i$ \\
		1.0 & 1.4 & $3.74510 - 0.651884 i$ & $3.77410 - 0.474190 i$ \\
		\hhline{====}
	\end{tabular}
	\caption{The quasinormal frequencies $\omega_0$ of the "non-RN" black holes, with $l=6$ and $n=0$. Here, $\omega_0$-Prony-non-RN denotes the quasinormal frequencies obtained using the finite difference method, while $\omega_0$-WKB-non-RN represents those calculated via the third-order WKB approximation.}
	\label{tab:omega_0_II}
\end{table}

Since the WKB method is an approximate approach, it is observed that the values of $\omega$ obtained from the two methods are not strictly identical. However, the results from both methods are in close agreement, indicating a high level of reliability in our findings. Notably, the imaginary parts of all quasinormal frequencies $\omega$ are negative, which confirms the stability of our black hole solutions.

\section{The Charge Mass Ratio $Q/M$ of Near-extremal Black Holes}\label{section 6}

For charged black holes, the extremal case is of particular interest. In this section, we investigate the properties of hairy charged black hole solution in the extremal limit.

\begin{figure}[H]	
	\centering
	\includegraphics[width=0.75\textwidth]{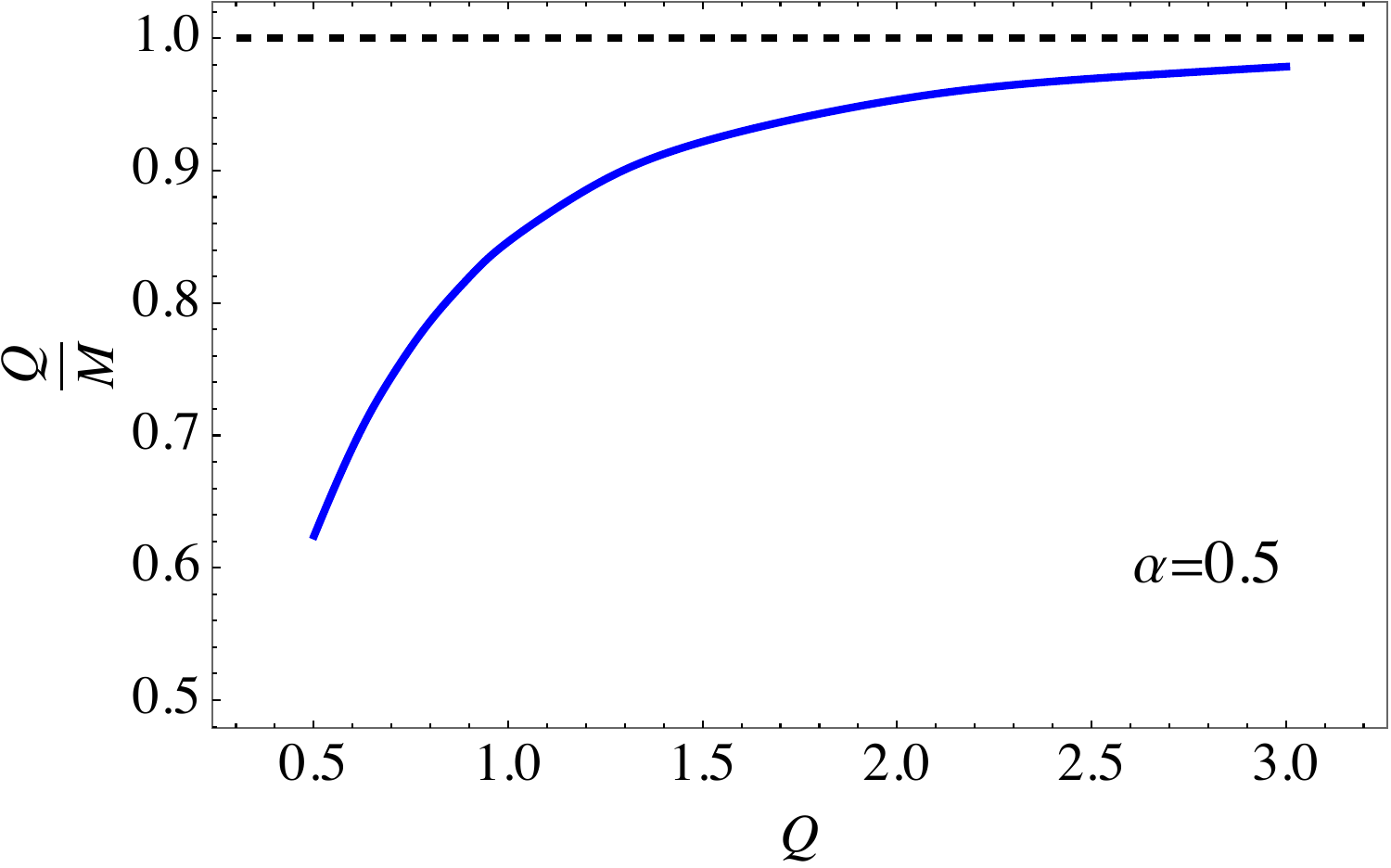}
	\caption{The charge to mass ratio $Q/M$ of RN-like black hole for $\alpha = 0.5 $ in the near-extremal limit.}
	\label{Fig8001}
\end{figure}

As is pointed out in the previous section, there are two branches of charged black hole solutions in Einstein-Weyl-Maxwell theory. One is totally different from RN black hole,  the temperature of this branch of charged black hole is far from zero and has no chance to achieve extremal limit, thus we shall not consider this branch of charged solutions. Instead, we shall focus on the other branch of charged black hole which is closer to RN black hole.  The exact extremal limit of charged black hole is hard to obtain through numerical method, instead we start from a non-extremal charged black hole with fixed charge and then change the initial conditions to approach the extremal limit.  In the case of $\alpha = 0.5$ and $Q=0.5$, we can obtain the near extremal charged black hole with temperature to be order of $10^{-4}$. Extracting the mass of the near extremal charged black hole, we can obtain the charge-to-mass ratio \(Q/M \), which is less than 1, the value of extremal RN black hole. Various values of charge are studied, and the ratio of charge over mass as a function of charge Q is plotted in Figure~\ref{Fig8001}.
It is obvious that charge-to-mass ratio of the charged black hole in Einstein-Weyl-Maxwell theory is always smaller than 1, the value of extremal RN black hole. It is worth pointing out that the value of charge-to-mass ratio approaches 1 as charge is increasing. It is consistent with the pattern we observe in the previous section, when charge is large the numerical charged black hole approaches the RN black hole.

\begin{figure}[H]	
	\centering
	\includegraphics[width=0.75\textwidth]{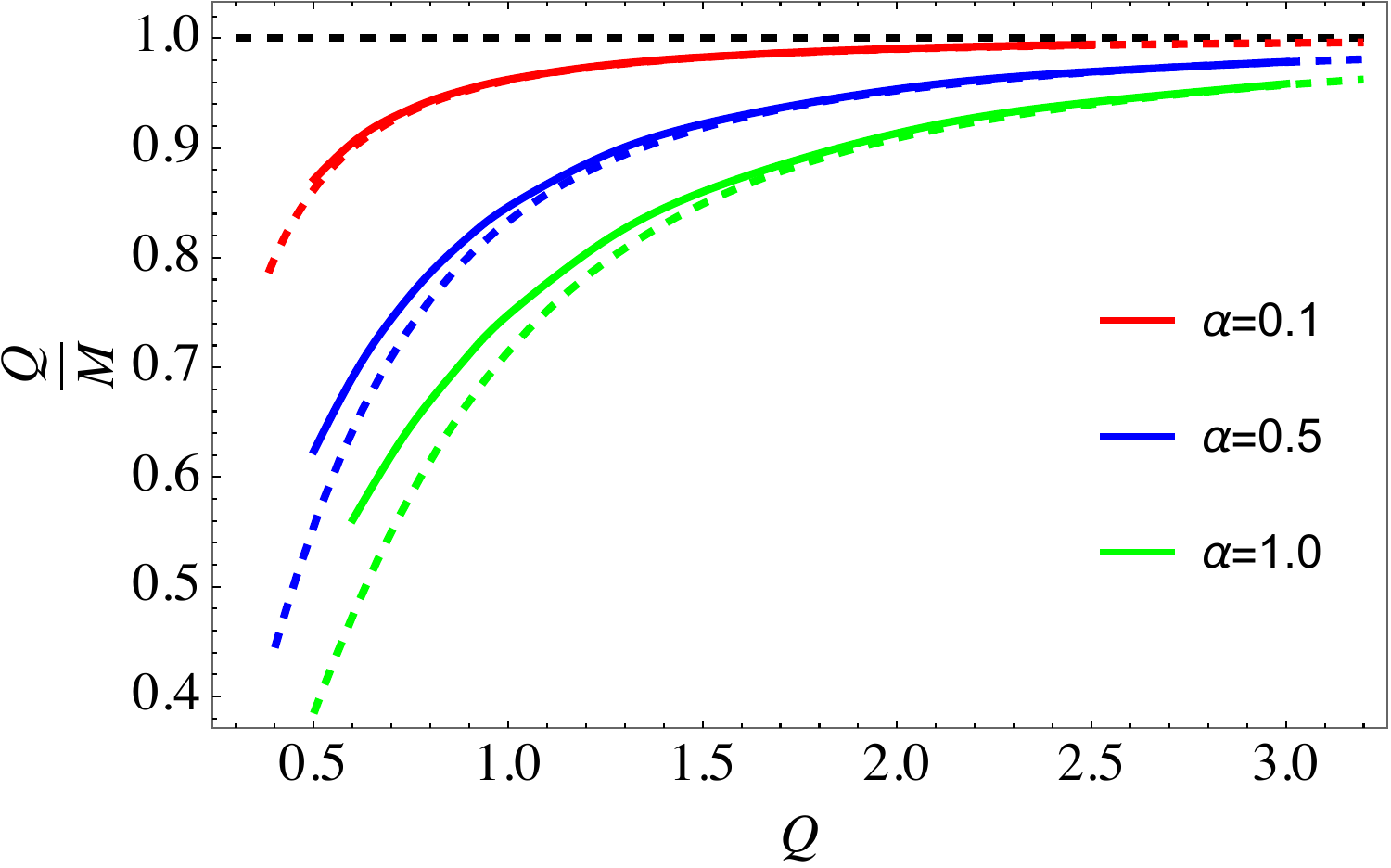}
	\caption{The charge-to-mass ratio $Q/M$ of the RN-like black hole is depicted for various coupling constants $\alpha$. The solid lines correspond to our numerical calculations, while the dashed lines indicate the results from perturbative calculations.}
	\label{Fig8003}
\end{figure}

In the literature, higher derivative terms were usually studied as quantum effects to Einstein-Maxwell theory, their corrections to charge-to-mass ratio of extremal RN black hole were calculated perturbatively~\cite{Kats:2006xp}. In~\cite{Kats:2006xp}, modifications up to four derivative terms were considered 
\begin{equation}
	\label{eq8001}
	\begin{aligned}
		S=\int d^4x&\sqrt{-g}\left(\frac{R}{2\kappa^2}-\frac{1}{4}F_{\mu\nu}F^{\mu\nu}+c_1R^2+c_2R_{\mu\nu}R^{\mu\nu}+c_3R_{\mu\nu\rho\sigma}R^{\mu\nu\rho\sigma}+\right.\\
		&+c_4RF_{\mu\nu}F^{\mu\nu}+c_5R^{\mu\nu}F_{\mu\rho}F_\nu^\rho+c_6R^{\mu\nu\rho\sigma}F_{\mu\nu}F_{\rho\sigma}+c_7(F_{\mu\nu}F^{\mu\nu})^2\\
		&+c_8\left(\nabla_\mu F_{\rho\sigma}\right)(\nabla^\mu F^{\rho\sigma})+c_9\left(\nabla_\mu F_{\rho\sigma}\right)(\nabla^\rho F^{\mu\sigma})).
	\end{aligned}
\end{equation}
The correction to mass-charge relation of extremal black holes is given by
\begin{equation}
	\label{eq8002}
	\frac\kappa{\sqrt{2}}\frac M{|Q|}=1-\frac2{5q^2}\left(2c_2+8c_3+\frac{2c_5}{\kappa^2}+\frac{2c_6}{\kappa^2}+\frac{8c_7}{\kappa^4}-\frac{2c_8}{\kappa^2}-\frac{c_9}{\kappa^2}\right).
\end{equation}
The theory turns to the Einstein-Weyl-Maxwell theory by setting the coefficients to be
\begin{equation}
	\label{eq8003}
	\begin{aligned}
		&\kappa={\sqrt{2}}, \qquad c_{1}=-\frac{\alpha}{12}, \qquad c_{2}=\frac{\alpha}{2}, \qquad c_{3}=-\frac{\alpha}{4},\\
		&c_{4}=c_{5}=c_{6}=c_{7}=c_{8}=c_{9}=0.
	\end{aligned}
\end{equation}
And the charge to mass ratio of extremal black hole turns out to be
\begin{equation}
	\label{eq8004}
	\frac{Q}{M}=1-\frac{2}{5}\frac{\alpha}{Q^2}.
\end{equation}
In the Einstein-Weyl-Maxwell theory, the coupling constant $\alpha$ should be positive to admit numerical black holes. Thus, the result above shows that the charge-to-mass ratio is smaller than $1$, which is consistent with our numerical result. In order to make a more comprehensive analysis  on the role of higher derivative Weyl term, we further  calculate the ratio of charge over mass for small  $\alpha =0.1$ and large $\alpha=1$, and the results are shown in Figure~\ref{Fig8003}. In Figure~\ref{Fig8003}, we also include the results of perturbative calculation which are denoted by dashed lines. 

It can be seen from the figure that our numerical results are close to but a slightly different from the perturbative results. One expected feature is that our numerical results are closer to perturbative results as $\alpha$ decreases, and the two almost coincide with each other when $\alpha =0.1$. Another feature is that charge-to-mass ratio of numerical solution and perturbative calculation approach each other as charge $Q$ increases. This feature is universal for various $\alpha$, and all curves approach the RN extremal black hole when $Q$ is large enough.  This feature is also consistent with our previous observation that one of the numerical charged solution is getting close to RN black hole as $Q$ is increasing. The reason for this phenomena is that the effect of higher curvature terms could be neglected when the charge is large, thus the system goes back to Einstein-Maxwell theory. 

Finally, it is worth pointing out that we consider the Einstein-Weyl-Maxwell theory as a classical theory, rather than taking the Weyl square term as quantum effect, the coefficient $\alpha$ of Weyl square term can be arbitrary and don't need to be small. Under this viewpoint, our results show that there exist charged black holes whose charge-to-mass ratio can be smaller than that of RN black hole. In the Weak Gravity Conjecture, we lower the bound of charge-to-mass ratio to be below 1.

\section{Summary and Discussion}\label{section 7}

In the context of Einstein-Weyl theory, it was previously noted in~\cite{Lu:2015cqa} that there are two distinct branches of static, spherically symmetric black hole solutions. One branch corresponds to the well-known Schwarzschild black hole, while the other is referred to as the non-Schwarzschild black hole. Subsequently, this framework was extended to include charged black holes by incorporating the Maxwell field into the Einstein-Weyl theory, as demonstrated in~\cite{Lin:2016jjl,Wu:2019uvq}. Within the Einstein-Weyl-Maxwell theory, it was claimed that there exist two branches of charged black hole solutions, with one branch being a generalization of the Schwarzschild solution and the other a generalization of the non-Schwarzschild solution.
However, upon conducting a thorough analysis, we have found that this classification of the two branches of charged black holes is not accurate. The relationship between the charged and neutral solutions is not one-to-one. Instead, one of the charged solutions incorporates elements of both the Schwarzschild and non-Schwarzschild solutions, as clearly illustrated in Figure~\ref{Fig30041}.

Despite the fact that the Einstein-Weyl-Maxwell theory does not admit the Reissner-Nordström (RN) black hole as a solution, we have nonetheless investigated the relationship between the RN solution and the hairy charged black holes for fixed charges. Our exploration has led to a significant discovery. We have identified that one of the hairy charged black holes is fundamentally different from the RN black hole, and we refer to this branch as the "non-RN" black holes. In contrast, the other hairy charged black hole is termed the "RN-like" black hole because of its similarity to the RN black hole. Notably, the RN-like black hole converges to the RN black hole for large charge Q.

The RN-like hairy charged solution possesses an extremal limit. We have investigated the charge-to-mass ratio of the RN-like black hole and discovered that its value is less than 1. We compared our findings with the perturbative results reported in the literature. Mathematically, our results are in close agreement with the perturbative calculations, and the deviation between our numerical results and the perturbative results diminishes as the coupling constant $\alpha$ decreases, which is consistent with expectations. As the charge Q increases, the charge-to-mass ratio approaches that of the extremal RN black hole.
Physically, however, there is a significant difference between the perturbative analysis and our full numerical consideration. In perturbative theory, higher derivative terms are treated as small quantum corrections, whereas we incorporate the higher derivative terms fully, allowing the coupling constant to be large. Our results establish a lower bound for the charge-to-mass ratio within the context of the Weak Gravity Conjecture.

\acknowledgments
We are grateful to Hong Lu for useful discussion. This work is supported in part by NSFC (National Natural Science Foundation of China) Grant No.~12075166.

\bibliographystyle{JHEP}

\bibliography{ref-Einstein-Weyl-Maxwell-Extremal-1}

\providecommand{\href}[2]{#2}\begingroup\raggedright\begin{thebibliography}{10}

\bibitem{LIGOScientific:2017vwq}
{\scshape LIGO Scientific, Virgo} collaboration, B.~P. Abbott et~al.,
  \emph{{GW170817: Observation of Gravitational Waves from a Binary Neutron
  Star Inspiral}},
  \href{http://dx.doi.org/10.1103/PhysRevLett.119.161101}{\emph{Phys. Rev.
  Lett.} {\bf 119} (2017) 161101}, [\href{http://arxiv.org/abs/1710.05832}{{\tt
  1710.05832}}].

\bibitem{LIGOScientific:2017zic}
{\scshape LIGO Scientific, Virgo, Fermi-GBM, INTEGRAL} collaboration, B.~P.
  Abbott et~al., \emph{{Gravitational Waves and Gamma-rays from a Binary
  Neutron Star Merger: GW170817 and GRB 170817A}},
  \href{http://dx.doi.org/10.3847/2041-8213/aa920c}{\emph{Astrophys. J. Lett.}
  {\bf 848} (2017) L13}, [\href{http://arxiv.org/abs/1710.05834}{{\tt
  1710.05834}}].

\bibitem{EventHorizonTelescope:2019dse}
{\scshape Event Horizon Telescope} collaboration, K.~Akiyama et~al.,
  \emph{{First M87 Event Horizon Telescope Results. I. The Shadow of the
  Supermassive Black Hole}},
  \href{http://dx.doi.org/10.3847/2041-8213/ab0ec7}{\emph{Astrophys. J. Lett.}
  {\bf 875} (2019) L1}, [\href{http://arxiv.org/abs/1906.11238}{{\tt
  1906.11238}}].

\bibitem{EventHorizonTelescope:2019uob}
{\scshape Event Horizon Telescope} collaboration, K.~Akiyama et~al.,
  \emph{{First M87 Event Horizon Telescope Results. II. Array and
  Instrumentation}},
  \href{http://dx.doi.org/10.3847/2041-8213/ab0c96}{\emph{Astrophys. J. Lett.}
  {\bf 875} (2019) L2}, [\href{http://arxiv.org/abs/1906.11239}{{\tt
  1906.11239}}].

\bibitem{EventHorizonTelescope:2019jan}
{\scshape Event Horizon Telescope} collaboration, K.~Akiyama et~al.,
  \emph{{First M87 Event Horizon Telescope Results. III. Data Processing and
  Calibration}},
  \href{http://dx.doi.org/10.3847/2041-8213/ab0c57}{\emph{Astrophys. J. Lett.}
  {\bf 875} (2019) L3}, [\href{http://arxiv.org/abs/1906.11240}{{\tt
  1906.11240}}].

\bibitem{EventHorizonTelescope:2019ths}
{\scshape Event Horizon Telescope} collaboration, K.~Akiyama et~al.,
  \emph{{First M87 Event Horizon Telescope Results. IV. Imaging the Central
  Supermassive Black Hole}},
  \href{http://dx.doi.org/10.3847/2041-8213/ab0e85}{\emph{Astrophys. J. Lett.}
  {\bf 875} (2019) L4}, [\href{http://arxiv.org/abs/1906.11241}{{\tt
  1906.11241}}].

\bibitem{EventHorizonTelescope:2019pgp}
{\scshape Event Horizon Telescope} collaboration, K.~Akiyama et~al.,
  \emph{{First M87 Event Horizon Telescope Results. V. Physical Origin of the
  Asymmetric Ring}},
  \href{http://dx.doi.org/10.3847/2041-8213/ab0f43}{\emph{Astrophys. J. Lett.}
  {\bf 875} (2019) L5}, [\href{http://arxiv.org/abs/1906.11242}{{\tt
  1906.11242}}].

\bibitem{EventHorizonTelescope:2019ggy}
{\scshape Event Horizon Telescope} collaboration, K.~Akiyama et~al.,
  \emph{{First M87 Event Horizon Telescope Results. VI. The Shadow and Mass of
  the Central Black Hole}},
  \href{http://dx.doi.org/10.3847/2041-8213/ab1141}{\emph{Astrophys. J. Lett.}
  {\bf 875} (2019) L6}, [\href{http://arxiv.org/abs/1906.11243}{{\tt
  1906.11243}}].

\bibitem{Stelle:1976gc}
K.~S. Stelle, \emph{{Renormalization of Higher Derivative Quantum Gravity}},
  \href{http://dx.doi.org/10.1103/PhysRevD.16.953}{\emph{Phys. Rev. D} {\bf 16}
  (1977) 953--969}.

\bibitem{Li:2008dq}
W.~Li, W.~Song and A.~Strominger, \emph{{Chiral Gravity in Three Dimensions}},
  \href{http://dx.doi.org/10.1088/1126-6708/2008/04/082}{\emph{JHEP} {\bf 04}
  (2008) 082}, [\href{http://arxiv.org/abs/0801.4566}{{\tt 0801.4566}}].

\bibitem{Deser:1981wh}
S.~Deser, R.~Jackiw and S.~Templeton, \emph{{Topologically Massive Gauge
  Theories}},
  \href{http://dx.doi.org/10.1016/0003-4916(82)90164-6}{\emph{Annals Phys.}
  {\bf 140} (1982) 372--411}.

\bibitem{Bergshoeff:2009aq}
E.~A. Bergshoeff, O.~Hohm and P.~K. Townsend, \emph{{More on Massive 3D
  Gravity}}, \href{http://dx.doi.org/10.1103/PhysRevD.79.124042}{\emph{Phys.
  Rev. D} {\bf 79} (2009) 124042}, [\href{http://arxiv.org/abs/0905.1259}{{\tt
  0905.1259}}].

\bibitem{Lu:2011zk}
H.~Lu and C.~N. Pope, \emph{{Critical Gravity in Four Dimensions}},
  \href{http://dx.doi.org/10.1103/PhysRevLett.106.181302}{\emph{Phys. Rev.
  Lett.} {\bf 106} (2011) 181302}, [\href{http://arxiv.org/abs/1101.1971}{{\tt
  1101.1971}}].

\bibitem{Deser:2011xc}
S.~Deser, H.~Liu, H.~Lu, C.~N. Pope, T.~C. Sisman and B.~Tekin, \emph{{Critical
  Points of D-Dimensional Extended Gravities}},
  \href{http://dx.doi.org/10.1103/PhysRevD.83.061502}{\emph{Phys. Rev. D} {\bf
  83} (2011) 061502}, [\href{http://arxiv.org/abs/1101.4009}{{\tt 1101.4009}}].

\bibitem{Lu:2015cqa}
H.~Lu, A.~Perkins, C.~N. Pope and K.~S. Stelle, \emph{{Black Holes in
  Higher-Derivative Gravity}},
  \href{http://dx.doi.org/10.1103/PhysRevLett.114.171601}{\emph{Phys. Rev.
  Lett.} {\bf 114} (2015) 171601}, [\href{http://arxiv.org/abs/1502.01028}{{\tt
  1502.01028}}].

\bibitem{Lin:2016jjl}
K.~Lin, A.~B. Pavan, G.~Flores-Hidalgo and E.~Abdalla, \emph{{New Electrically
  Charged Black Hole in Higher Derivative Gravity}},
  \href{http://dx.doi.org/10.1007/s13538-017-0505-0}{\emph{Braz. J. Phys.} {\bf
  47} (2017) 419--425}, [\href{http://arxiv.org/abs/1605.04562}{{\tt
  1605.04562}}].

\bibitem{Wu:2019uvq}
C.~Wu, D.-C. Zou and M.~Zhang, \emph{{Charged black holes in the
  Einstein-Maxwell-Weyl gravity}},
  \href{http://dx.doi.org/10.1016/j.nuclphysb.2020.114942}{\emph{Nucl. Phys. B}
  {\bf 952} (2020) 114942}, [\href{http://arxiv.org/abs/1904.10193}{{\tt
  1904.10193}}].

\bibitem{Smilga_2014}
A.~V. Smilga, \emph{Supersymmetric field theory with benign ghosts},
  \href{http://dx.doi.org/10.1088/1751-8113/47/5/052001}{\emph{Journal of
  Physics A: Mathematical and Theoretical} {\bf 47} (Jan., 2014) 052001}.

\bibitem{L__2017}
H.~Lü, A.~Perkins, C.~Pope and K.~Stelle, \emph{Lichnerowicz modes and black
  hole families in ricci quadratic gravity},
  \href{http://dx.doi.org/10.1103/physrevd.96.046006}{\emph{Physical Review D}
  {\bf 96} (Aug., 2017) }.

\bibitem{Wald1983}
R.~M. Wald, \emph{The thermodynamics of black holes},
  \href{http://dx.doi.org/10.1007/BF01206013}{\emph{Communications in
  Mathematical Physics} {\bf 87} (1983) 259--276}.

\bibitem{IyerWald1994}
V.~Iyer and R.~M. Wald, \emph{Some properties of noether charge and a proposal
  for dynamical black hole entropy},
  \href{http://dx.doi.org/10.1103/PhysRevD.50.846}{\emph{Physical Review D}
  {\bf 50} (1994) 846--864}.

\bibitem{Gundlach:1993tp}
C.~Gundlach, R.~H. Price and J.~Pullin, \emph{{Late time behavior of stellar
  collapse and explosions: 1. Linearized perturbations}},
  \href{http://dx.doi.org/10.1103/PhysRevD.49.883}{\emph{Phys. Rev. D} {\bf 49}
  (1994) 883--889}, [\href{http://arxiv.org/abs/gr-qc/9307009}{{\tt
  gr-qc/9307009}}].

\bibitem{Iyer:1986np}
S.~Iyer and C.~M. Will, \emph{{Black Hole Normal Modes: A {WKB} Approach. 1.
  Foundations and Application of a Higher Order {WKB} Analysis of Potential
  Barrier Scattering}},
  \href{http://dx.doi.org/10.1103/PhysRevD.35.3621}{\emph{Phys. Rev. D} {\bf
  35} (1987) 3621}.

\bibitem{Kats:2006xp}
Y.~Kats, L.~Motl and M.~Padi, \emph{{Higher-order corrections to mass-charge
  relation of extremal black holes}},
  \href{http://dx.doi.org/10.1088/1126-6708/2007/12/068}{\emph{JHEP} {\bf 12}
  (2007) 068}, [\href{http://arxiv.org/abs/hep-th/0606100}{{\tt
  hep-th/0606100}}].

\end{thebibliography}\endgroup

\end{document}